\pdfoutput=1
\documentclass[12pt,a4paper]{article}
\usepackage{ifthen} % for conditional statements
\newboolean{pdflatex}
\setboolean{pdflatex}{true} % False for eps figures 

\newboolean{articletitles}
\setboolean{articletitles}{true} % False removes titles in references

\newboolean{uprightparticles}
\setboolean{uprightparticles}{false} %True for upright particle symbols

\def\paperauthors{LHCb collaboration} % Leave as is for PAPER, CONF and FIGURE
\def\paperasciititle{Measurement of CP violation in B02DD and Bs2DsDs decays} % Set ASCII title here !! MAKE sure it's only ASCII characters !! 
\def\papertitle{Measurement of \CP violation\\ in \decay{\Bd}{\Dp\Dm} and \decay{\Bs}{\Dsp\Dsm} decays} % Latex formatted title
\def\paperkeywords{{High Energy Physics}, {LHCb}} % Comma separated list
\def\papercopyright{\the\year\ CERN for the benefit of the LHCb collaboration} % new since 9/Apr/2018
\def\paperlicence{CC BY 4.0 licence}
\def\paperlicenceurl{https://creativecommons.org/licenses/by/4.0/}

\newif\ifEnableSectionTOCLinks
\EnableSectionTOCLinksfalse % deactivated

\usepackage[top=1in, bottom=1.25in, left=1in, right=1in]{geometry}

\usepackage{microtype}
\usepackage{lineno}  % for line numbering during review
\usepackage{xspace} % To avoid problems with missing or double spaces after
\usepackage{caption} %these three command get the figure and table captions automatically small

\usepackage{graphicx}  % to include figures (can also use other packages)
\usepackage{color}
\usepackage{colortbl}
\graphicspath{{./figs/}} % Make Latex search fig subdir for figures
\usepackage{amsmath} % Adds a large collection of math symbols
\usepackage{amssymb}
\usepackage{amsfonts}
\usepackage{upgreek} % Adds in support for greek letters in roman typeset

\newcommand*\patchAmsMathEnvironmentForLineno[1]{%
\expandafter\let\csname old#1\expandafter\endcsname\csname #1\endcsname
\expandafter\let\csname oldend#1\expandafter\endcsname\csname
end#1\endcsname
 \renewenvironment{#1}%
   {\linenomath\csname old#1\endcsname}%
   {\csname oldend#1\endcsname\endlinenomath}%
}
\newcommand*\patchBothAmsMathEnvironmentsForLineno[1]{%
  \patchAmsMathEnvironmentForLineno{#1}%
  \patchAmsMathEnvironmentForLineno{#1*}%
}
\AtBeginDocument{%
\patchBothAmsMathEnvironmentsForLineno{equation}%
\patchBothAmsMathEnvironmentsForLineno{align}%
\patchBothAmsMathEnvironmentsForLineno{flalign}%
\patchBothAmsMathEnvironmentsForLineno{alignat}%
\patchBothAmsMathEnvironmentsForLineno{gather}%
\patchBothAmsMathEnvironmentsForLineno{multline}%
\patchBothAmsMathEnvironmentsForLineno{eqnarray}%
}

\usepackage{hyperref}
\usepackage{hyperxmp}
\hypersetup{pdftex,
pdfauthor={\paperauthors},
pdftitle={\paperasciititle},
pdfkeywords={\paperkeywords},
pdfcopyright={Copyright (C) \papercopyright},
pdflicenseurl={\paperlicenceurl}}

\usepackage[colorinlistoftodos,textsize=scriptsize]{todonotes}

\usepackage[bottom,flushmargin,hang,multiple]{footmisc}

\usepackage[all]{hypcap} % Internal hyperlinks to floats.

\usepackage{xspace} 
\usepackage{upgreek}

\def\lhcb   {\mbox{LHCb}\xspace}

\def\babar  {\mbox{BaBar}\xspace}
\def\belle  {\mbox{Belle}\xspace}

\def\MagUp {\mbox{\em Mag\kern -0.05em Up}\xspace}

\ifthenelse{\boolean{uprightparticles}}%
{

 \def\Ppi         {\ensuremath{\uppi}\xspace}

 \def\Pphi        {\ensuremath{\upphi}\xspace}

 \def\Ppsi        {\ensuremath{\uppsi}\xspace}

 \def\PDelta      {\ensuremath{\Delta}\xspace}                 
 \def\PXi         {\ensuremath{\Xi}\xspace}                 
 \def\PLambda     {\ensuremath{\Lambda}\xspace}                 
 \def\PSigma      {\ensuremath{\Sigma}\xspace}                 
 \def\POmega      {\ensuremath{\Omega}\xspace}                 
 \def\PUpsilon    {\ensuremath{\Upsilon}\xspace}
 \let\oldPi\Pi
 \def\PPi         {\ensuremath{\oldPi}\xspace}

 \def\PB      {\ensuremath{\mathrm{B}}\xspace}                 
 \def\PD      {\ensuremath{\mathrm{D}}\xspace}                 
 \def\PJ      {\ensuremath{\mathrm{J}}\xspace}                 
 \def\PK      {\ensuremath{\mathrm{K}}\xspace}                 
 \def\Pb      {\ensuremath{\mathrm{b}}\xspace}                 
 \def\Pc      {\ensuremath{\mathrm{c}}\xspace}                 
 \def\Pd      {\ensuremath{\mathrm{d}}\xspace}

 \def\Ph      {\ensuremath{\mathrm{h}}\xspace}                 
 \def\Pp      {\ensuremath{\mathrm{p}}\xspace}                 
 \def\Pq      {\ensuremath{\mathrm{q}}\xspace}                 
 \def\Ps      {\ensuremath{\mathrm{s}}\xspace}

 \def\thebaroffset{0.0em}
}
{

 \def\Ppi         {\ensuremath{\pi}\xspace}

 \def\Pphi        {\ensuremath{\phi}\xspace}

 \def\Ppsi        {\ensuremath{\psi}\xspace}                 
                  
 \mathchardef\PDelta="7101
 \mathchardef\PXi="7104
 \mathchardef\PLambda="7103
 \mathchardef\PSigma="7106
 \mathchardef\POmega="710A
 \mathchardef\PUpsilon="7107
 \mathchardef\PPi="7105
 \def\PB      {\ensuremath{B}\xspace}                 
 \def\PD      {\ensuremath{D}\xspace}                 
 \def\PJ      {\ensuremath{J}\xspace}                 
 \def\PK      {\ensuremath{K}\xspace}                 
 \def\Pb      {\ensuremath{b}\xspace}                 
 \def\Pc      {\ensuremath{c}\xspace}                 
 \def\Pd      {\ensuremath{d}\xspace}

 \def\Ph      {\ensuremath{h}\xspace}                 
 \def\Pp      {\ensuremath{p}\xspace}                 
 \def\Pq      {\ensuremath{q}\xspace}                 
 \def\Ps      {\ensuremath{s}\xspace}

 \def\thebaroffset{0.18em}
}
\newcommand{\offsetoverline}[2][\thebaroffset]{\kern #1\overline{\kern -#1 #2}}%

\makeatletter
\ifcase \@ptsize \relax% 10pt
  \newcommand{\miniscule}{\@setfontsize\miniscule{4}{5}}% \tiny: 5/6
\or% 11pt
  \newcommand{\miniscule}{\@setfontsize\miniscule{5}{6}}% \tiny: 6/7
\or% 12pt
  \newcommand{\miniscule}{\@setfontsize\miniscule{5}{6}}% \tiny: 6/7
\fi
\makeatother

\DeclareRobustCommand{\optbar}[1]{\shortstack{{\miniscule (\rule[.5ex]{1.25em}{.18mm})}
  \\ [-.7ex] $#1$}}

\def\quark     {{\ensuremath{\Pq}}\xspace}

\def\dquark    {{\ensuremath{\Pd}}\xspace}

\def\squark    {{\ensuremath{\Ps}}\xspace}

\def\cquark    {{\ensuremath{\Pc}}\xspace}

\def\bquark    {{\ensuremath{\Pb}}\xspace}
\def\bquarkbar {{\ensuremath{\overline \bquark}}\xspace}

\def\hadron {{\ensuremath{\Ph}}\xspace}
\def\pion   {{\ensuremath{\Ppi}}\xspace}

\def\pip    {{\ensuremath{\pion^+}}\xspace}
\def\pim    {{\ensuremath{\pion^-}}\xspace}

\def\kaon    {{\ensuremath{\PK}}\xspace}
\def\KorKbar {\kern \thebaroffset\optbar{\kern -\thebaroffset \PK}{}\xspace}

\def\Kp      {{\ensuremath{\kaon^+}}\xspace}
\def\Km      {{\ensuremath{\kaon^-}}\xspace}

\def\KS      {{\ensuremath{\kaon^0_{\mathrm{S}}}}\xspace}

\newcommand{\phiz}{\ensuremath{\Pphi}\xspace}

\def\D       {{\ensuremath{\PD}}\xspace}

\def\DorDbar {\kern \thebaroffset\optbar{\kern -\thebaroffset \PD}\xspace}
\def\Dz      {{\ensuremath{\D^0}}\xspace}

\def\Dp      {{\ensuremath{\D^+}}\xspace}
\def\Dm      {{\ensuremath{\D^-}}\xspace}
\def\Dpm     {{\ensuremath{\D^\pm}}\xspace}
\def\Dmp     {{\ensuremath{\D^\mp}}\xspace}
\def\DpDm    {\ensuremath{\Dp {\kern -0.16em \Dm}}\xspace}

\def\Dstarpm {{\ensuremath{\D^{*\pm}}}\xspace}

\def\Dsp     {{\ensuremath{\D^+_\squark}}\xspace}
\def\Dsm     {{\ensuremath{\D^-_\squark}}\xspace}
\def\Dspm    {{\ensuremath{\D^{\pm}_\squark}}\xspace}

\def\DporDsp {{\ensuremath{\D_{(\squark)}^+}}\xspace}
\def\DmorDsm {{\ensuremath{\D{}_{(\squark)}^-}}\xspace}

\def\B       {{\ensuremath{\PB}}\xspace}
\def\Bbar    {{\ensuremath{\offsetoverline{\PB}}}\xspace}

\def\BorBbar {\kern \thebaroffset\optbar{\kern -\thebaroffset \PB}\xspace}

\def\Bzb     {{\ensuremath{\Bbar{}^0}}\xspace}
\def\Bd      {{\ensuremath{\B^0}}\xspace}
\def\Bdb     {{\ensuremath{\Bbar{}^0}}\xspace}
\def\BdorBdbar {\kern \thebaroffset\optbar{\kern -\thebaroffset \Bd}\xspace}

\def\Bs      {{\ensuremath{\B^0_\squark}}\xspace}
\def\Bsb     {{\ensuremath{\Bbar{}^0_\squark}}\xspace}
\def\BsorBsbar {\kern \thebaroffset\optbar{\kern -\thebaroffset \Bs}\xspace}

\def\Bds     {{\ensuremath{\B_{(\squark)}^0}}\xspace}
\def\Bdsb    {{\ensuremath{\Bbar{}_{(\squark)}^0}}\xspace}
\def\BdorBs  {\Bds}
\def\BdorBsbar  {\Bdsb}

\def\jpsi     {{\ensuremath{{\PJ\mskip -3mu/\mskip -2mu\Ppsi}}}\xspace}

\def\Y#1S{\ensuremath{\PUpsilon{(#1S)}}\xspace}

\def\proton      {{\ensuremath{\Pp}}\xspace}

\def\Lz          {{\ensuremath{\PLambda}}\xspace}

\def\LorLbar     {\kern \thebaroffset\optbar{\kern -\thebaroffset \PLambda}\xspace}

\def\Lc          {{\ensuremath{\Lz^+_\cquark}}\xspace}

\newcommand{\decay}[2]{\ensuremath{#1\!\to #2}\xspace} 

\def\to                 {\ensuremath{\rightarrow}\xspace}

\newcommand{\tauBs}{{\ensuremath{\tau_{\Bs}}}\xspace}
\newcommand{\tauBd}{{\ensuremath{\tau_{\Bd}}}\xspace}

\def\CP                {{\ensuremath{C\!P}}\xspace}

\newcommand{\dm}{{\ensuremath{\Delta m}}\xspace}
\newcommand{\dms}{{\ensuremath{\Delta m_{\squark}}}\xspace}
\newcommand{\dmd}{{\ensuremath{\Delta m_{\dquark}}}\xspace}
\newcommand{\DG}{{\ensuremath{\Delta\Gamma}}\xspace}
\newcommand{\DGs}{{\ensuremath{\Delta\Gamma_{\squark}}}\xspace}
\newcommand{\DGd}{{\ensuremath{\Delta\Gamma_{\dquark}}}\xspace}

\newcommand{\phid}{{\ensuremath{\phi_{\dquark}}}\xspace}

\newcommand{\phis}{{\ensuremath{\phi_{\squark}}}\xspace}

\newcommand{\mistag}{\ensuremath{\omega}\xspace}

\def\AT#1     {\ensuremath{A_{\mathrm{T}}^{#1}}\xspace}           % 2

\def\C#1      {\ensuremath{\mathcal{C}_{#1}}\xspace}                       % 9
\def\Cp#1     {\ensuremath{\mathcal{C}_{#1}^{'}}\xspace}                    % 7
\def\Ceff#1   {\ensuremath{\mathcal{C}_{#1}^{\mathrm{(eff)}}}\xspace}        % 9  
\def\Cpeff#1  {\ensuremath{\mathcal{C}_{#1}^{'\mathrm{(eff)}}}\xspace}       % 7
\def\Ope#1    {\ensuremath{\mathcal{O}_{#1}}\xspace}                       % 2
\def\Opep#1   {\ensuremath{\mathcal{O}_{#1}^{'}}\xspace}                    % 7

\newcommand{\nospaceunit}[1]{\ensuremath{\text{#1}}}       
\newcommand{\aunit}[1]{\ensuremath{\text{\,#1}}}       
\newcommand{\tev}{\aunit{Te\kern -0.1em V}\xspace}
\newcommand{\gev}{\aunit{Ge\kern -0.1em V}\xspace}
\newcommand{\mev}{\aunit{Me\kern -0.1em V}\xspace}
\newcommand{\kev}{\aunit{ke\kern -0.1em V}\xspace}
\newcommand{\ev}{\aunit{e\kern -0.1em V}\xspace}
 
\newcommand{\mevc}{\ensuremath{\aunit{Me\kern -0.1em V\!/}c}\xspace}
\newcommand{\gevc}{\ensuremath{\aunit{Ge\kern -0.1em V\!/}c}\xspace}
\newcommand{\mevcc}{\ensuremath{\aunit{Me\kern -0.1em V\!/}c^2}\xspace}
\newcommand{\gevcc}{\ensuremath{\aunit{Ge\kern -0.1em V\!/}c^2}\xspace}
\def\mm   {\aunit{mm}\xspace}

\def\mum  {\ensuremath{\,\upmu\nospaceunit{m}}\xspace}

\def\fb   {\ensuremath{\aunit{fb}}\xspace}
\def\invfb   {\ensuremath{\fb^{-1}}\xspace}

\def\ps   {\ensuremath{\aunit{ps}}\xspace}
\def\fs   {\aunit{fs}}

\def\invps{\ensuremath{\ps^{-1}}\xspace}

\newcommand{\stat}{\aunit{(stat)}\xspace}
\newcommand{\syst}{\aunit{(syst)}\xspace}

\newcommand{\chisq}{\ensuremath{\chi^2}\xspace}

\newcommand{\chisqip}{\ensuremath{\chi^2_{\text{IP}}}\xspace}

\def\deriv {\ensuremath{\mathrm{d}}}

\def\gsim{{~\raise.15em\hbox{$>$}\kern-.85em
          \lower.35em\hbox{$\sim$}~}\xspace}
\def\lsim{{~\raise.15em\hbox{$<$}\kern-.85em
          \lower.35em\hbox{$\sim$}~}\xspace}

\def\sPlot{\mbox{\em sPlot}\xspace}

\def\pt         {\ensuremath{p_{\mathrm{T}}}\xspace}

\def\ptot       {\ensuremath{p}\xspace}

\def\rad{\aunit{rad}\xspace}

\def\evtgen     {\mbox{\textsc{EvtGen}}\xspace}

\def\geant      {\mbox{\textsc{Geant4}}\xspace}

\def\photos     {\mbox{\textsc{Photos}}\xspace}

\def\pythia     {\mbox{\textsc{Pythia}}\xspace}

\def\tell1  {TELL1\xspace}
\def\ukl1   {UKL1\xspace}

\newcommand{\lhcborcid}[1]{\href{https://orcid.org/#1}{\hspace*{0.1em}\raisebox{-0.45ex}{\includegraphics[width=1em]{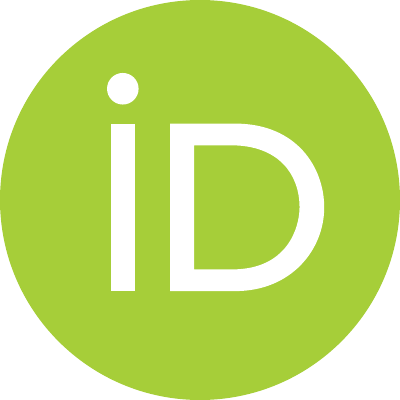}}}}

\hypersetup{
  colorlinks   = true, %Colours links instead of ugly boxes
  urlcolor     = blue, %Colour for external hyperlinks
  linkcolor    = blue, %Colour of internal links
  citecolor    = red   %Colour of citations
}

\ifEnableSectionTOCLinks
    \usepackage[explicit]{titlesec} % to change headings
    
    \let\oldcontentsline\contentsline
    \renewcommand

    \titleformat{\section}{\normalfont\Large\bf}{\hyperlink{tocsection.\thesection}{{\thesection} \parbox[t]{\dimexpr\textwidth-1pc}{#1}}}{1pc}{}

    \titleformat{\subsection}{\normalfont\bf}{\hyperlink{tocsubsection.\thesubsection}{{\thesubsection} \parbox[t]{\dimexpr\textwidth-1pc}{#1}}}{1pc}{}

    \titleformat{name=\section,numberless}[display]{}{}{0pt}{\normalfont\Huge\bfseries #1}
\fi

\usepackage{cite} % Allows for ranges in citations
\usepackage{mciteplus}
\usepackage{cleveref}
\crefname{chapter}{Ch.\@}{Chs.\@}
\crefname{section}{Sec.\@}{Secs.\@}
\crefname{subsection}{Sec.\@}{Secs.\@}
\crefname{appendix}{Appendix\@}{Appendices\@}
\crefname{figure}{Fig.\@}{Figs.\@}
\crefname{table}{Table\@}{Tables\@}
\crefname{equation}{Eq.\@}{Eqs.\@}

\usepackage{subcaption} % enables subfigures

\def\SResultRunTwo{-0.552 \pm 0.100\stat \pm 0.010\syst}
\def\CResultRunTwo{0.128 \pm0.103\stat \pm 0.010\syst}
\def\SResultCombined{-0.549 \pm 0.085\stat \pm 0.015\syst}
\def\CResultCombined{0.162 \pm0.088\stat \pm 0.009\syst}

\def\phisResultRunTwo{-0.086 \pm 0.106 \stat \pm 0.028\syst \rad}
\def\lambdaResultRunTwo{1.145 \pm 0.126\stat \pm 0.031\syst}
\def\phisResultCombined{-0.055 \pm 0.090 \stat \pm 0.021\syst \rad}
\def\lambdaResultCombined{1.054 \pm 0.099 \stat \pm 0.020\syst}

\usepackage{longtable} % only for template; not usually to be used in PAPERs

\begin{document}

%%%%%%%%%%%%%%%%%%%%%%%%%
%%%%% Title     %%%%%%%%%
%%%%%%%%%%%%%%%%%%%%%%%%%
\renewcommand{\thefootnote}{\fnsymbol{footnote}}
\setcounter{footnote}{1}

% %%%%%%% CHOOSE TITLE PAGE--------
%\onecolumn
%\input{title-LHCb-INT}
%\input{title-LHCb-ANA}
%\input{title-LHCb-CONF}
%\input{title-LHCb-FIGURE}
% ===============================================================================
% Purpose: LHCb-PAPER journal paper title page template
% Author: 
% Created on: 2010-09-25
% ===============================================================================

%%%%%%%%%%%%%%%%%%%%%%%%%
%%%%%  TITLE PAGE  %%%%%%
%%%%%%%%%%%%%%%%%%%%%%%%%
\begin{titlepage}
\pagenumbering{roman}

% Header ---------------------------------------------------
\vspace*{-1.5cm}
\centerline{\large EUROPEAN ORGANIZATION FOR NUCLEAR RESEARCH (CERN)}
\vspace*{1.5cm}
\noindent
\begin{tabular*}{\linewidth}{lc@{\extracolsep{\fill}}r@{\extracolsep{0pt}}}
\ifthenelse{\boolean{pdflatex}}% Logo format choice
{\vspace*{-1.5cm}\mbox{\!\!\!\includegraphics[width=.14\textwidth]{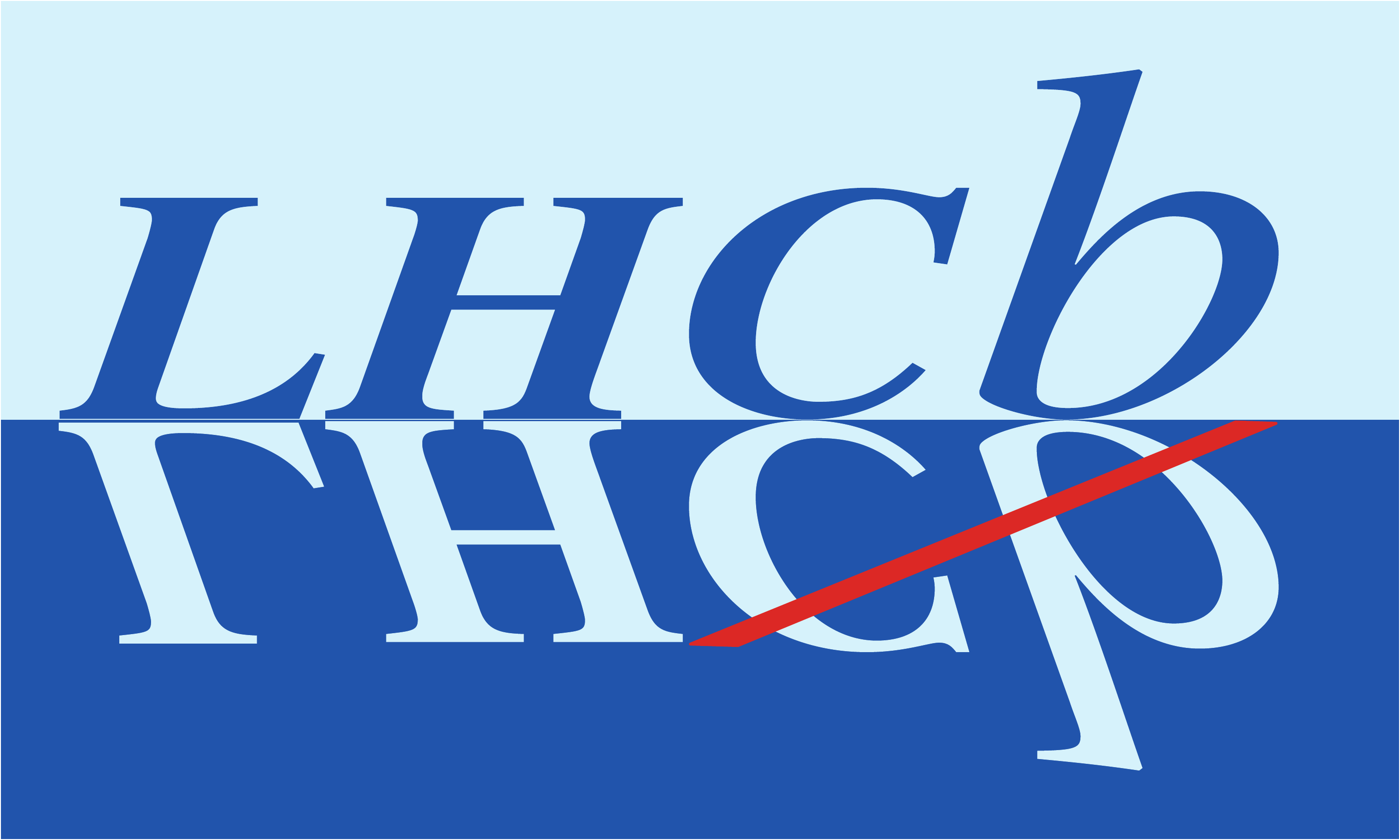}} & &}%
{\vspace*{-1.2cm}\mbox{\!\!\!\includegraphics[width=.12\textwidth]{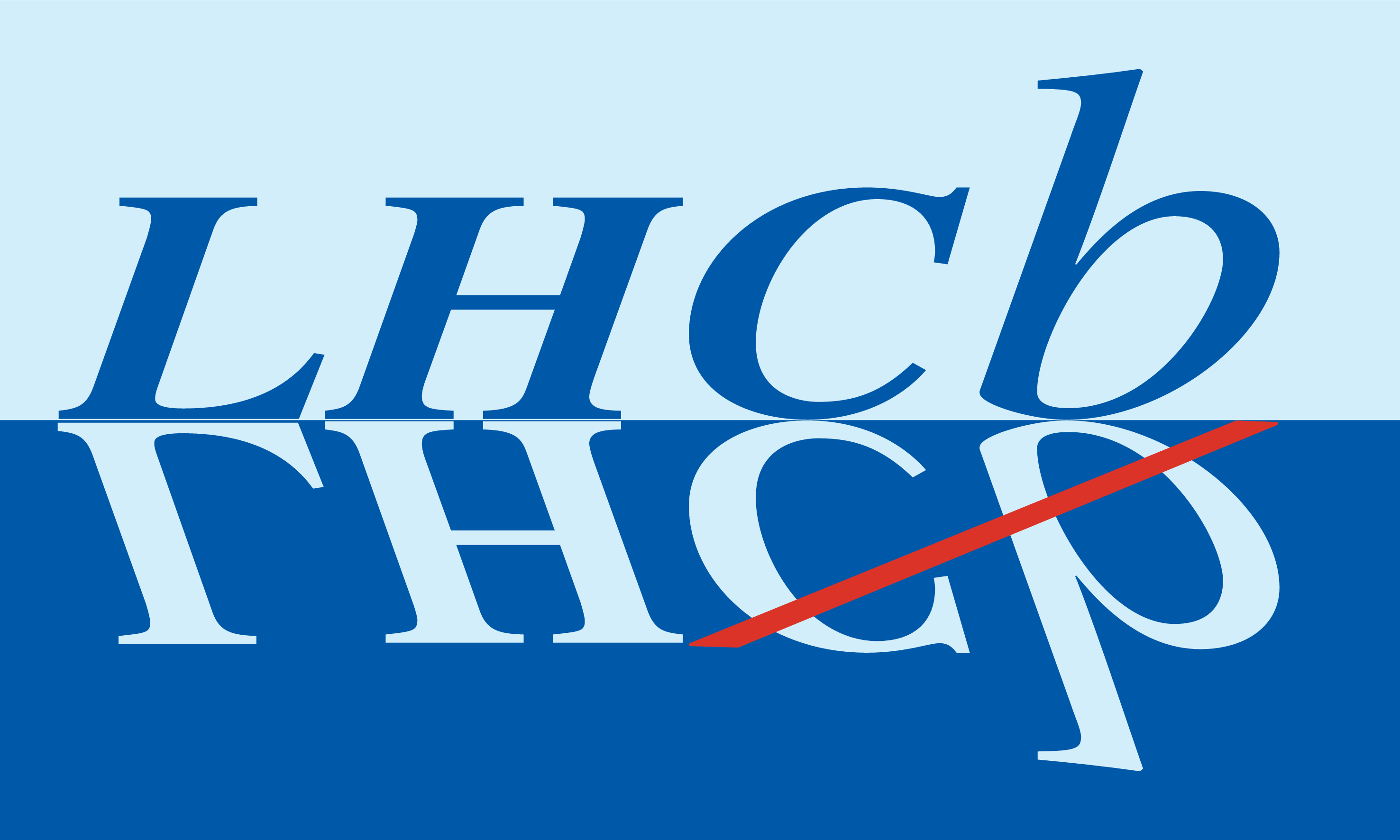}} & &}%
\\
 & & CERN-EP-2024-217 \\  % ID 
 & & LHCb-PAPER-2024-027  \\  % ID 
 & & January 15, 2025 \\ % Date - Can also hardwire e.g.: 23 March 2010
 & & \\
% not in paper \hline
\end{tabular*}

\vspace*{3.0cm}

% Title --------------------------------------------------
{\normalfont\bfseries\boldmath\huge
\begin{center}
% DO NOT EDIT HERE. Instead edit macro in main.tex to keep metadata correct
  \papertitle 
\end{center}
}

\vspace*{1.0cm}

% Authors -------------------------------------------------
\begin{center}
%In the footnote, replace 'paper' by 'Letter' in case of submission to PRL or PLB 
% Edit macro in main.tex to keep metadata correct
\paperauthors\footnote{Authors are listed at the end of this paper.}
\end{center}

\vspace{\fill}

% Abstract -----------------------------------------------
\begin{abstract}
  \noindent
  A time-dependent, flavour-tagged measurement of \CP violation is performed with
  \decay{\Bd}{\Dp\Dm} and \decay{\Bs}{\Dsp\Dsm} decays, using data collected by the \lhcb
  detector in proton-proton collisions at a centre-of-mass energy of 13\tev corresponding to an integrated luminosity of
  6\invfb.
  In \decay{\Bd}{\Dp\Dm} decays the \CP-violation parameters are measured to be
  \begin{align}
    S_{\Dp\Dm} & = \SResultRunTwo, \nonumber           \\
    C_{\Dp\Dm} & = \phantom{-}\CResultRunTwo. \nonumber
  \end{align}
  In \decay{\Bs}{\Dsp\Dsm} decays the \CP-violating parameter formulation in terms of \phis and $|\lambda|$ results in
\begin{align}
    \phis                & = \phisResultRunTwo, \nonumber              \\
    |\lambda_{\Dsp\Dsm}| & = \phantom{-}\lambdaResultRunTwo. \nonumber
\end{align}
These results represent the most precise single measurement of the \CP-violation parameters in their respective channels. For
the first time in a single measurement, \CP symmetry is observed to be violated in \decay{\Bd}{\Dp\Dm} decays with a significance exceeding six standard deviations.
\end{abstract}

\vspace*{1.0cm}

\begin{center}
  % Submitted to
  Published in JHEP 01 (2025) 061
  % JHEP /
  % Phys.~Rev.~D /
  % Phys.~Rev.~Lett. /
%  Phys.~Lett.~B /
  % Eur.~Phys.~J.~C /
  %  Nucl.~Phys.~B /
  % Chin.~Phys.~C /
  % Nature~Physics /
  % sciPost~Physics /
  % J. Instr. /
  % Instruments 
\end{center}

\vspace{\fill}

{\footnotesize 
% Edit macro in main.tex to keep metadata correct
\centerline{\copyright~\papercopyright. \href{\paperlicenceurl}{\paperlicence}.}}
\vspace*{2mm}

\end{titlepage}

%%%%%%%%%%%%%%%%%%%%%%%%%%%%%%%%
%%%%%  EOD OF TITLE PAGE  %%%%%%
%%%%%%%%%%%%%%%%%%%%%%%%%%%%%%%%

%  empty page follows the title page ----
\newpage
\setcounter{page}{2}
\mbox{~}
%\newpage
%
%% Author List ----------------------------
%%  You need to get a new author list!
%\input{LHCb_authorlist.tex}
%
%The author list for journal publications is provided by the Membership Committee shortly after 'approval to go to paper' has been given.
%%It will be made available on the page
%%\verb!http://www.physik.uzh.ch/~strauman/forMemCo/LHCb-PAPER-XXXX-XXX/! .
%It will be sent to you by email shortly after a paper number has beens assigned.
%The author list should be included already at first circulation, 
%to allow new members of the collaboration to verify whether they have been included correctly.
%Occasionally a misspelled name is corrected or associated institutions become full members.
%In that case, a new author list will be sent to you.
%In case line numbering doesn't work well after including the authorlist, try moving the \verb!\bigskip! after the last author to a separate line.
%
%
%The authorship for Conference Reports should be ``The LHCb
%  collaboration'', with a footnote giving the name(s) of the contact
%  author(s), but without the full list of collaboration names.

%\twocolumn
% %%%%%%%%%%%%% ---------

\renewcommand{\thefootnote}{\arabic{footnote}}
\setcounter{footnote}{0}

%%%%%%%%%%%%%%%%%%%%%%%%%%%%%%%%
%%%%%  Table of Content   %%%%%%
%%%%%%%%%%%%%%%%%%%%%%%%%%%%%%%%
%%%% Uncomment if desired
%\tableofcontents

\cleardoublepage

%%%%%%%%%%%%%%%%%%%%%%%%%
%%%%% Main text %%%%%%%%%
%%%%%%%%%%%%%%%%%%%%%%%%%

\pagestyle{plain} % restore page numbers for the main text
\setcounter{page}{1}
\pagenumbering{arabic}

%% Uncomment during review phase. 
%% Comment before a final submission.
%\linenumbers

%% This is the main body
%% It is useful to have a single file so comments are not missed in overleaf.
\section{Introduction}
\label{sec:intro}

Measurements of \CP violation in \BdorBs mesons play a crucial role in the search for physics beyond the Standard Model (SM).
With the increase in experimental precision, control over hadronic matrix elements becomes more important, which
is a major challenge in most decay modes.
In decays of beauty mesons to two charmed mesons \decay{\B}{\D\D}, this can be achieved by employing U-spin flavour symmetry
and constraining the hadronic contributions by relating different \CP-violation and branching fraction measurements~\cite{Fleischer_2007,Fleischer_1999,Jung_2015,Bel_2015,Davies2024}.

\begin{figure}[b]
    \centering
    \includegraphics[width=0.49\linewidth]{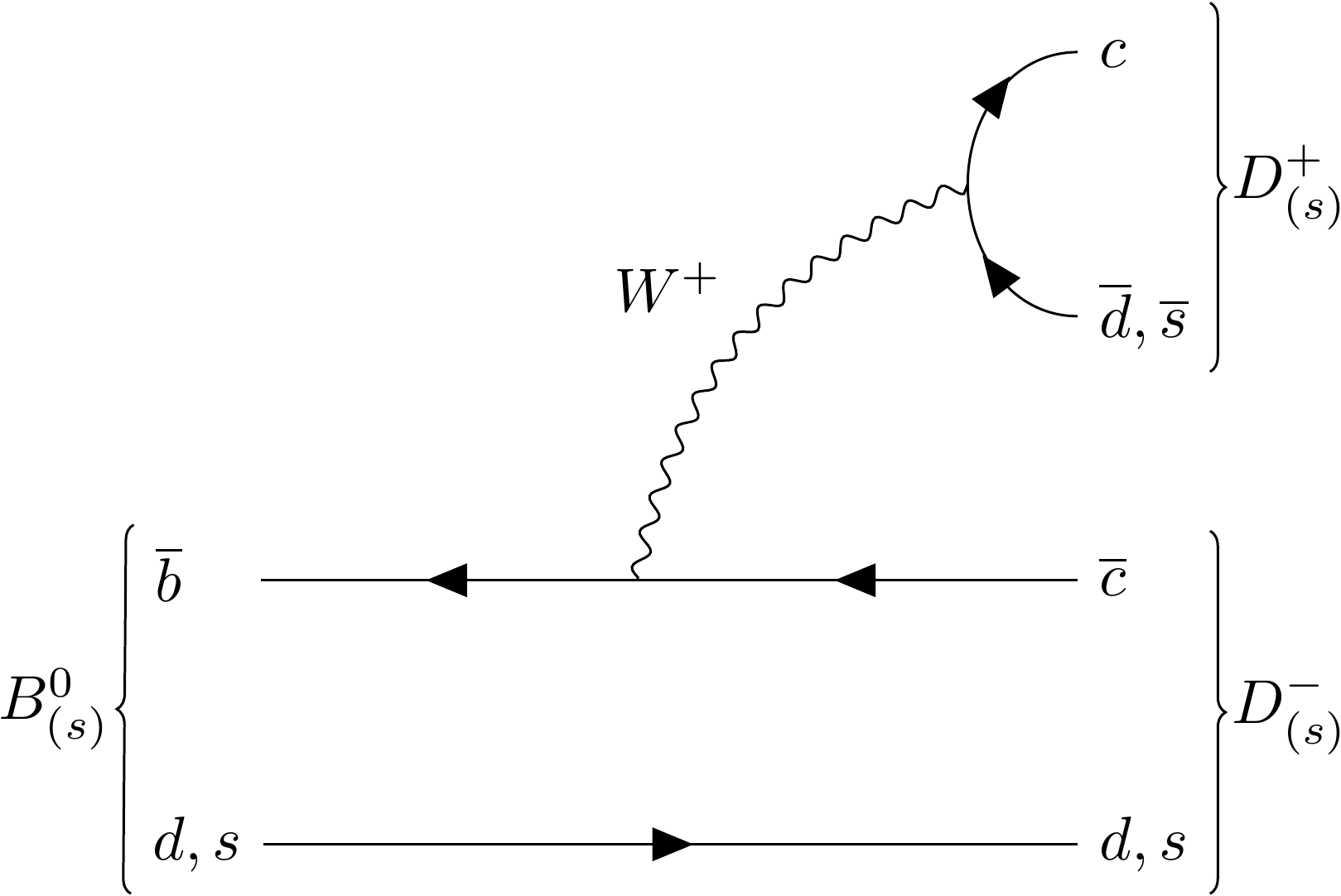}
    \includegraphics[width=0.49\linewidth]{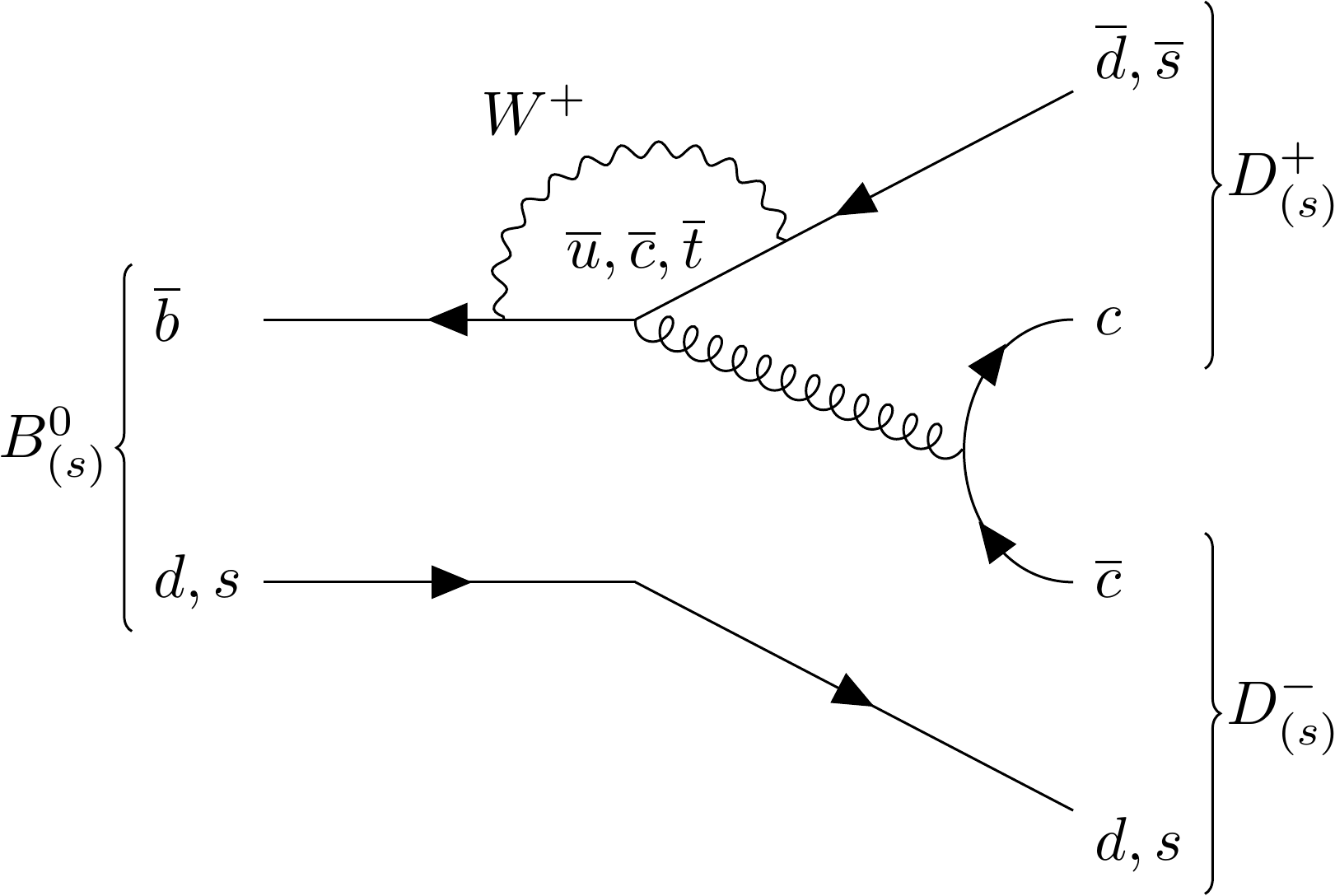}
    
    \vspace{0.5cm}
    
    \includegraphics[width=0.49\linewidth]{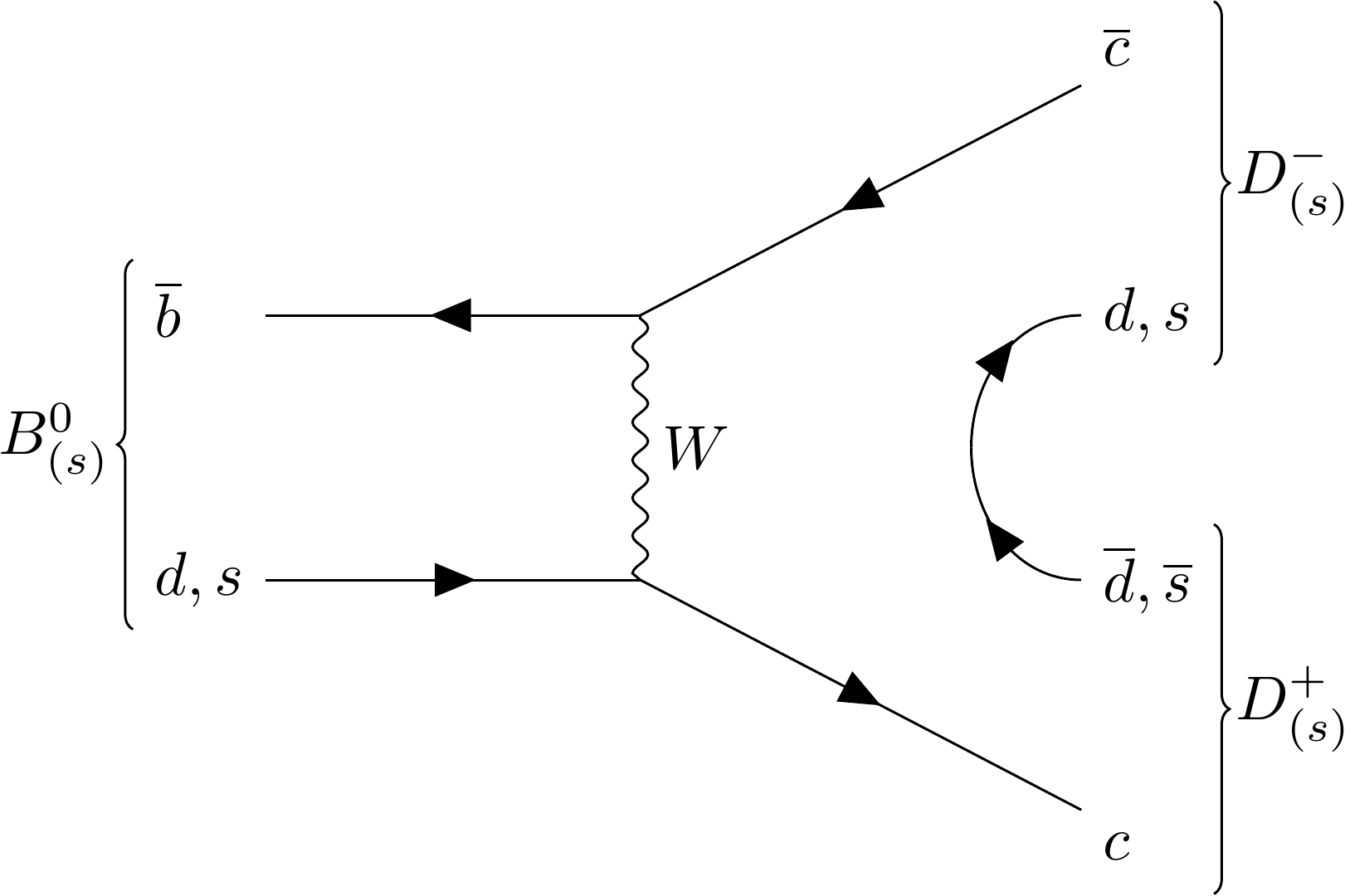}
    \includegraphics[width=0.49\linewidth]{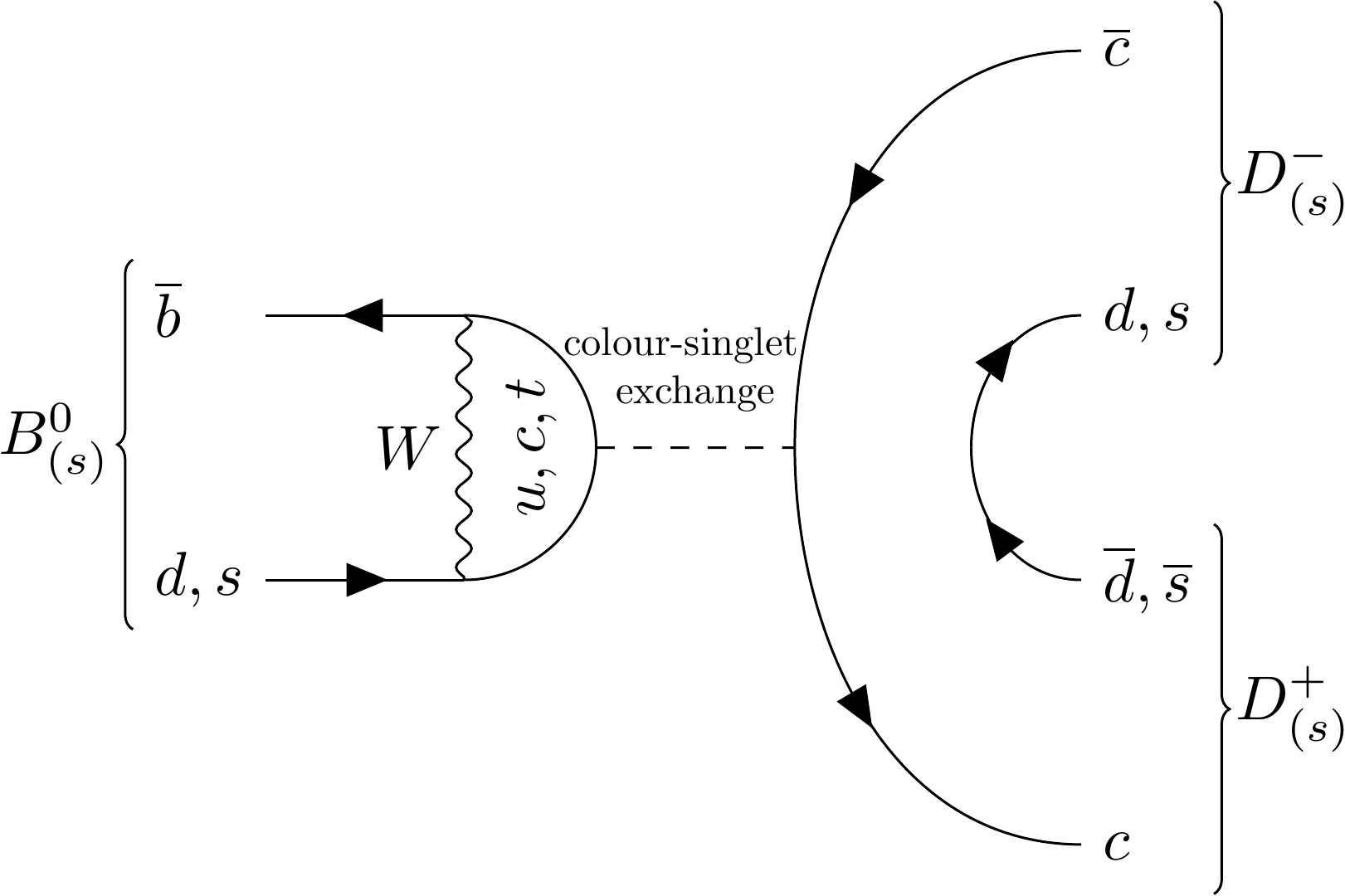}
    \caption{Dominant Feynman diagrams contributing to the $B^{0} \rightarrow D^{+} D^{-}$ and \mbox{$B_{s}^{0} \rightarrow D_{s}^{+} D_{s}^{-}$} decays.
        The (top left) tree-level, (bottom left) exchange, (top right) penguin and (bottom right) penguin annihilation diagrams are shown.}
    \label{fig:feynman}
\end{figure}

The \decay{\B}{\D\D} system gives access to a variety of interesting observables
that probe elements of the Cabibbo--Kobayashi--Maskawa (CKM) quark-mixing matrix~\cite{Cabibbo:1963yz,Kobayashi:1973fv}.
In \decay{\Bd}{\Dp\Dm} and \decay{\Bs}{\Dsp\Dsm} decays, the \CP-violating weak phases $\beta$ and $\beta_s$ can be measured, respectively.
The phases arise in the interference between the \Bd--\Bdb (\Bs--\Bsb) mixing and the tree-level decay amplitudes to the \Dp\Dm (\Dsp\Dsm) final state, leading to time-dependent \CP asymmetries.
The decays can also proceed through several other diagrams, as shown in \cref{fig:feynman}. The \CP asymmetries may arise from both SM contributions and new physics effects, if present.

In \decay{\Bd}{\Dp\Dm} and \decay{\Bs}{\Dsp\Dsm} decays, the same final state is accessible from both \BdorBs and \BdorBsbar states.
The partial decay rate as a function of the decay time $t$ is given by
\begin{align}
    \label{eq:decayrate}
    \frac{\deriv\Gamma(t,d)}{\deriv t} \propto
    e^{-t/\tau_{\BdorBs}} \left( \cosh{\frac{\DG_{\quark} t}{2}}
    + D_f\sinh{\frac{\DG_{\quark} t}{2}}
    + d\,C_f\cos{\dm_{\quark} t}
    - d\,S_f\sin{\dm_{\quark} t} \right),
\end{align}
where $\DG_{\quark} = \Gamma_{\quark\text{L}} - \Gamma_{\quark\text{H}}$ and $\dm_{\quark} = m_{\quark\text{H}} - m_{\quark\text{L}}$ are the decay-width difference and mass difference of the heavy and light \Bd ($\quark=\dquark$) or \Bs ($\quark=\squark$) mass eigenstates,
$\tau_{\BdorBs}$ is the mean lifetime of the \BdorBs meson and the tag $d$ represents the flavour at production taking the value
$+1$ for a \BdorBs meson and $-1$ for a \BdorBsbar meson.
The \CP-violation parameters are defined as
\begin{equation}
    \begin{gathered}
        \label{eq:cpparameters}
        D_f = -\frac{2|\lambda_f|\cos{\phi_{\quark}}}{1+|\lambda_f|^2},\ C_f = \frac{1-|\lambda_f|^2}{1+|\lambda_f|^2},\ S_f = -\frac{2|\lambda_f|\sin{\phi_{\quark}}}{1+|\lambda_f|^2}, \\
        \lambda_f = \frac{q}{p}\frac{\bar{A}_f}{A_f} \text{ and } \phi_{\quark} = -\arg\lambda_f,
    \end{gathered}
\end{equation}
where $A_f$ and $\bar{A}_f$  are the decay amplitudes of \BdorBs and \BdorBsbar to the common final state $f$ and the ratio $q/p$ describes mixing of the \BdorBs mesons.
The parameter $D_f$ cannot be measured in \Bd decays because, at the current experimental precision, \DGd
is compatible with zero. Thus, the decay rates for \decay{\Bd}{\Dp\Dm} can be simplified to
\begin{align}
    \label{eq:decayrate_b2dd}
    \frac{\deriv\Gamma(t,d)}{\deriv t} \propto
    e^{-t/\tauBd} \left(1
    + d\,C_{\Dp\Dm}\cos{\dmd t}
    - d\,S_{\Dp\Dm}\sin{\dmd t} \right).
\end{align}
If only tree-level contributions in \decay{\Bd}{\Dp\Dm} decays are considered, direct \CP violation vanishes resulting in $C_{\Dp\Dm}=0$ and
$S_{\Dp\Dm} = -\sin{\phid} = -\sin{2\beta}$.
This assumption is valid within the current experimental precision for \decay{\Bd}{\jpsi\KS} decays, where $\beta$ can be measured
with high precision as recently reported by \lhcb~\cite{LHCb-PAPER-2023-013}.
However, in \decay{\Bd}{\Dp\Dm} measurements the loop-mediated penguin contributions shown in \cref{fig:feynman} cannot be neglected and
an additional phase shift is measured via $\sin{(2\beta + \Delta\phid)} = - S_{\Dp\Dm}/\sqrt{1-C^{2}_{\Dp\Dm}}$.
This measurement enables higher-order corrections to the measurement of \phis in \mbox{\decay{\Bs}{\Dsp\Dsm}} decays to be constrained,
under the assumption of U-spin flavour symmetry.

Due to the similarities of the two decay channels, a parallel measurement of the \CP-violation parameters in \decay{\Bd}{\Dp\Dm}
and \decay{\Bs}{\Dsp\Dsm} decays is performed.
Both decays have been previously studied by \lhcb~\cite{LHCb-Paper-2016-037,LHCb-PAPER-2014-051}, while measurements of
the \mbox{\CP-violation} parameters in \decay{\Bd}{\Dp\Dm} decays have been performed by \babar~\cite{PhysRevD.79.032002} and \belle~\cite{PhysRevD.85.091106}.
The \belle result lies outside the physically allowed region and shows a small tension with the other measurements.

This analysis uses proton-proton (pp) collision data collected by the \lhcb experiment during the years 2015 to 2018 corresponding to an integrated luminosity of 6\invfb.
The \decay{\Bd}{\Dp\Dm} candidates are reconstructed through the decays \decay{\Dp}{\Km\pip\pip}
and \decay{\Dp}{\Km\Kp\pip}.\footnote{If not stated otherwise, charge-conjugated decays are implied.}
These decays have the highest branching fractions into charged kaons and pions.
Candidates where both \Dpm mesons decay via \decay{\Dp}{\Km\Kp\pip} are not considered due to the
smaller branching fraction of this mode.
Similarly, one of the \Dspm mesons from the \decay{\Bs}{\Dsp\Dsm} candidates is always reconstructed through the decay \decay{\Dsp}{\Km\Kp\pip}
and the other is reconstructed through the decays \decay{\Dsp}{\Km\Kp\pip}, \decay{\Dsp}{\pim\Kp\pip} or \decay{\Dsp}{\pim\pip\pip}.

Both signal channels and a dedicated \decay{\Bd}{\Dsp\Dm} control channel are selected by similar criteria with only minor differences as described in \cref{sec:selection}.
A mass fit is performed separately for each final state to statistically subtract the remaining background as described in \cref{sec:massfit}.
The knowledge of the initial flavour of the candidates is crucial for measurements of time-dependent asymmetries in neutral \B-meson decays.
In \cref{sec:tagging} the algorithms used to determine the initial flavour of the \BdorBs mesons are described.
The decay-time fit to measure the \CP-violation parameters is described in \cref{sec:decaytimefit} and the systematic uncertainties are discussed in \cref{sec:systematics}.
In \cref{sec:results} the results are presented from both this analysis and in combination with previous \lhcb measurements.

\section{Detector and simulation}
\label{sec:Detector}
The \lhcb detector~\cite{LHCb-DP-2008-001,LHCb-DP-2014-002} is a single-arm forward
spectrometer covering the \mbox{pseudorapidity} range $2<\eta <5$,
designed for the study of particles containing \bquark or \cquark
quarks. The detector includes a high-precision tracking system
consisting of a silicon-strip vertex detector surrounding the $pp$
interaction region~\cite{LHCb-DP-2014-001}, a large-area silicon-strip detector located
upstream of a dipole magnet with a bending power of about
$4{\mathrm{\,T\,m}}$, and three stations of silicon-strip detectors and straw
drift tubes~\cite{LHCb-DP-2017-001} placed downstream of the magnet.
The tracking system provides a measurement of the momentum, \ptot, of charged particles with
a relative uncertainty that varies from 0.5\% at low momentum to 1.0\% at 200\gevc.
The minimum distance of a track to a primary $pp$ collision vertex (PV), the impact parameter (IP),
is measured with a resolution of $(15+29/\pt)\mum$,
where \pt is the component of the momentum transverse to the beam, in\,\gevc.
Different types of charged hadrons are distinguished using information
from two ring-imaging Cherenkov detectors~\cite{LHCb-DP-2012-003}.
Photons, electrons and hadrons are identified by a calorimeter system consisting of
scintillating-pad and preshower detectors, an electromagnetic
% calorimeter
and a hadronic calorimeter. Muons are identified by a
system composed of alternating layers of iron and multiwire
proportional chambers~\cite{LHCb-DP-2012-002}.

Simulation is required to model the effects of the detector acceptance and the
imposed selection requirements.
Samples of signal decays are used to determine the parameterisation of the signal mass
distributions and decay-time resolution model.
In the simulation, $pp$ collisions are generated using
\pythia~\cite{Sjostrand:2007gs,*Sjostrand:2006za}
with a specific \lhcb configuration~\cite{LHCb-PROC-2010-056}.
Decays of unstable particles
are described by \evtgen~\cite{Lange:2001uf}, in which final-state
radiation is generated using \photos~\cite{davidson2015photos}.
The interaction of the generated particles with the detector, and its response,
are implemented using the \geant
toolkit~\cite{Allison:2006ve, *Agostinelli:2002hh} as described in
Ref.~\cite{LHCb-PROC-2011-006}.
The underlying $pp$ interaction is reused multiple times, with an independently generated signal decay for each~\cite{LHCb-DP-2018-004}.
To account for differences between the distributions of particle identification (PID) variables in simulation and data,
the PIDCalib package~\cite{LHCb-DP-2018-001} is used to reweight the distributions in the simulation.

\section{Selection}
\label{sec:selection}
% trigger
The online event selection is performed by a trigger~\cite{LHCb-DP-2012-004},
which consists of a hardware stage based on information from the calorimeter and muon
systems, followed by a software stage which applies a full event
reconstruction.
At the hardware trigger stage, events are required to have a muon with high \pt or a
hadron, photon or electron with high transverse energy in the calorimeters.
The software trigger requires a two-, three- or four-track
secondary vertex with a significant displacement from any primary
$pp$ interaction vertex. At least one charged particle
must have a transverse momentum $\pt > 1.6\gevc$ and be
inconsistent with originating from a PV.
A multivariate algorithm~\cite{BBDT,LHCb-PROC-2015-018} is used for
the identification of secondary vertices consistent with the decay
of a \bquark hadron.

% Stripping
In the offline selection, \Dpm and \Dspm candidates are reconstructed through their decays into the selected final-state particles, which are required to
satisfy loose selection criteria on their momentum, transverse momentum and PID variables, and be inconsistent with originating from any PV.
The \Dpm and \Dspm candidates should form vertices with a good fit quality and the scalar sum of transverse momenta of their three final-state particles should be greater than $1800\mevc$.
All possible combinations of tracks forming a common vertex should have a distance of closest approach smaller than $0.5\mm$.
The \BdorBs candidates are reconstructed from two \Dpm or \Dspm candidates with opposite charges that form a good-quality vertex.
The momentum vector of the \BdorBs candidates should point from the PV to the secondary vertex.
The scalar sum of the transverse momenta of all six final-state particles is required to be greater than $5000\mevc$.
% preselection
The invariant masses of the \Dpm and \Dspm candidates are required to be within a window of $\pm 45\mevcc$ around their known values~\cite{PDG2024}.
This requirement, of about $\pm4$ times the mass resolution, retains almost all candidates while separating the \Dpm from the \Dspm mass region.
To suppress single-charm decays of the form \decay{\BdorBs}{\DporDsp\hadron^+\hadron^-\hadron^-}, both \DporDsp candidates are required to
have a significant flight distance from the \BdorBs decay vertex ($>5\sigma$).

% Vetoes
In the reconstruction of the \DporDsp candidates, background contributions can arise from the misidentification of the final-state particles.
Misidentification from a pion, kaon or proton is considered.
The three-body invariant masses are recomputed to identify background decays from \Dp, \Dsp and \Lc states.
The masses for potential two-body background contributions arising from intermediate \phiz and \Dz decays are similarly computed.
These background sources are suppressed by PID requirements within the mass windows of the known particle masses.

% Veto BDT
A particularly challenging background arises from the misidentification between \mbox{$\decay{\Dp}{\Km\pip\pip}$} and \decay{\Dsp}{\Km\Kp\pip} decays.
The $\pip\leftrightarrow\Kp$ misidentification shifts the mass region of the reconstructed \Dp candidates to that of the \Dsp or vice versa.
In this case, a simple PID requirement does not provide the necessary rejection of
the particularly large background contribution from \decay{\Bd}{\Dsp\Dm} decays.
To distinguish between the two decays a boosted decision tree (BDT) algorithm is trained utilising the \texttt{xgboost} module from the \texttt{scikit-learn} package~\cite{Scikit-learn-paper}.
Simulated \Dp and \Dsp decays from the \decay{\Bd}{\Dp\Dm}, \decay{\Bd}{\Dsp\Dm} and \decay{\Bs}{\Dsp\Dsm} samples are used to train the BDT classifier.
A $k$-folding procedure with $k=5$ is used to avoid overtraining~\cite{10.1145/307400.307439}.
Various two- and three-body invariant masses, recomputed with different final-state particle hypotheses, are used in the training.
Additionally, the flight distance of the \DporDsp candidates, and the PID variables of those particles that are potentially misidentified, are used.
The requirements on the BDT-classifier output are chosen to suppress the \Dsp candidates in the \decay{\Bd}{\Dp\Dm} channel and
\Dp candidates in the \decay{\Bs}{\Dsp\Dsm} channel to negligible levels.
This is verified by applying the requirements to the simulated samples, which results in the rejection of more than $99\%$ of the respective candidates.

% combinatorial BDT
A second BDT classifier is trained to suppress combinatorial background.
As a signal proxy, all available simulated \decay{\Bd}{\Dp\Dm}, \decay{\Bd}{\Dsp\Dm} and \decay{\Bs}{\Dsp\Dsm} samples are used while
the background proxy is taken from the upper-mass sideband of the data, which is defined as $m_{\DporDsp\DmorDsm}>5600\mevcc$, beyond the \BdorBs-candidate mass fit region.
The variables used in the training are all transverse momenta of intermediate and final-state particles;
the flight distance and the difference in invariant mass from the known value~\cite{PDG2024} of the \DporDsp candidates;
the angle between the \DporDsp flight direction and each of the decay products;
the \chisqip of the \BdorBs and \DporDsp candidates,
which is the difference in the \chisq value of the PV fit with and without the particle being considered in the calculation.
Similar to the strategy used in Ref.~\cite{LHCb-PAPER-2016-027}, the requirement on the BDT-classifier output is chosen to minimise the uncertainties on the \CP-violation parameters.

% DecayTreeFitter
The invariant mass used in the mass fits is computed from a kinematic fit to the decay chain
with constraints on all charm-meson masses to improve the invariant-mass resolution of the \BdorBs candidates~\cite{Hulsbergen:2005pu}.
For calculation of the decay time, a constraint on the PV is used in the kinematic fit. To avoid correlations between the decay time and the invariant mass, no constraints on the charm-masses are used.

% final selection
Contributions from partially reconstructed backgrounds are reduced to negligible levels by restricting the invariant mass of the \Bd candidates to lie within the range 5240--5540\mevcc.
The decay-time range is chosen to be 0.3--10.3\ps, where the lower boundary is set to reduce background originating from the PV.
For \Bs candidates the same decay-time range is chosen, but the invariant-mass range is 5300--5600\mevcc.

After the selection, multiple candidates are found in about 1\% of the events.
Usually, these candidates differ in just one track or PID assignment.
Since it is very unlikely to find two genuine candidates in one event, only one of the candidates is chosen in a reproducible random way.

\section{Mass fit}
\label{sec:massfit}
An extended unbinned maximum-likelihood fit to the invariant mass of the \BdorBs candidates is performed to
extract per-event weights via the \sPlot technique~\cite{Pivk:2004ty}.
These weights are used in the decay-time fit to statistically subtract the background.
Pseudoexperiment studies indicate that any residual correlation between the decay time and the mass introduces no meaningful bias into the \CP-violation measurement.

% mass fits B02DD
The mass model in the \decay{\Bd}{\Dp\Dm} channel consists of a signal component and
two background components to model \decay{\Bs}{\Dp\Dm} decays and the combinatorial background.
A double-sided Hypatia probability density function (PDF)~\cite{Santos:2013gra} is used to model the signal component.
The shape parameters are determined by a fit to simulated \decay{\Bd}{\Dp\Dm} decays and fixed
in the fit to data, while the peak position and width of the distribution are allowed to vary.
The same model is used for the \decay{\Bs}{\Dp\Dm} component with a shift of the peak position by the known
mass difference between the \Bd and \Bs mesons~\cite{PDG2024}.
An exponential PDF is used to model the combinatorial background.
\begin{figure}[tb]
    \begin{center}
        \includegraphics[width=0.49\linewidth]{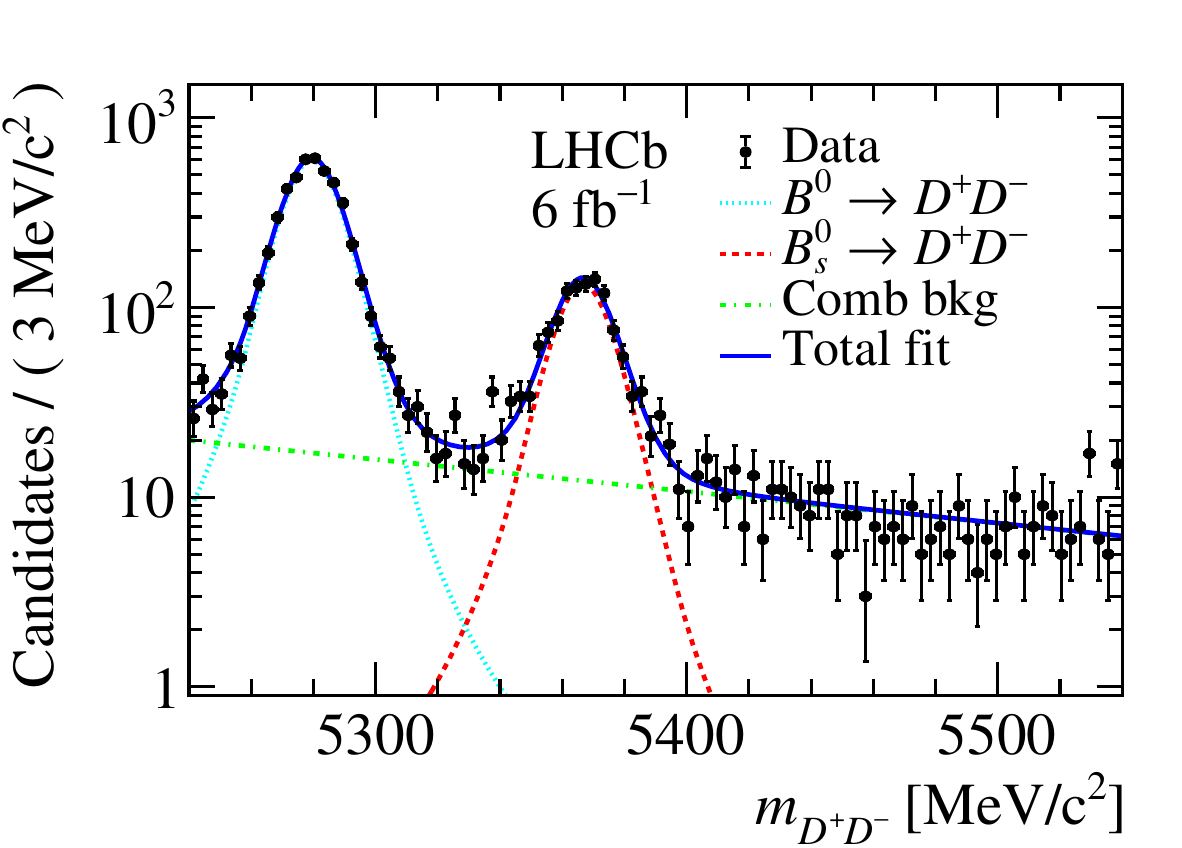}
        \includegraphics[width=0.49\linewidth]{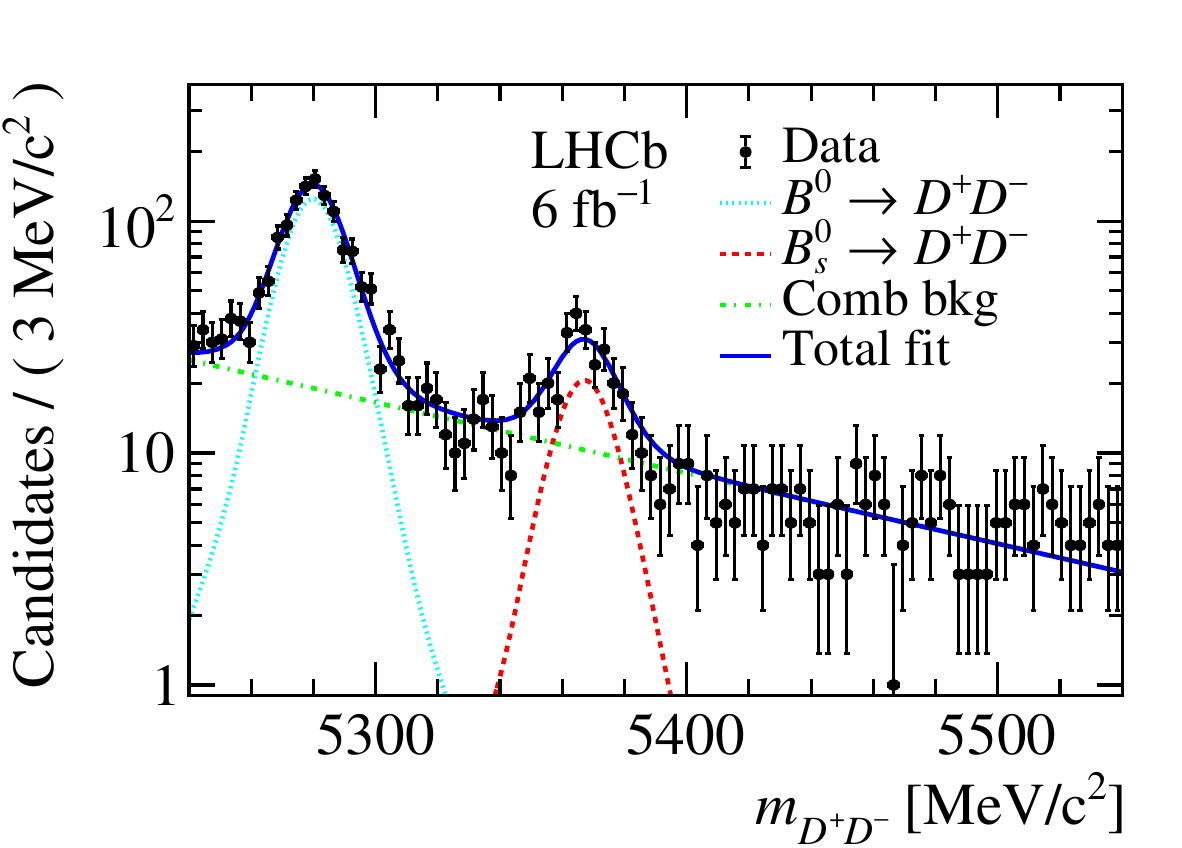}
        \vspace*{-0.5cm}
    \end{center}
    \caption{Invariant-mass distribution of $B^{0} \rightarrow D^{+} D^{-}$ decays.
        The data are shown as points and the full PDF is shown as a solid-blue line for
        (left) both $D^{\pm}$ candidates decaying through $D^{+} \rightarrow K^{-}\pi^{+}\pi^{+}$ and (right) one $D^{\pm}$ candidate decaying through $D^{+} \rightarrow K^{-}K^{+}\pi^{+}$.}
    \label{fig:massfits_btodd}
\end{figure}

\begin{figure}[!tb]
    \begin{center}
        \includegraphics[width=0.49\linewidth]{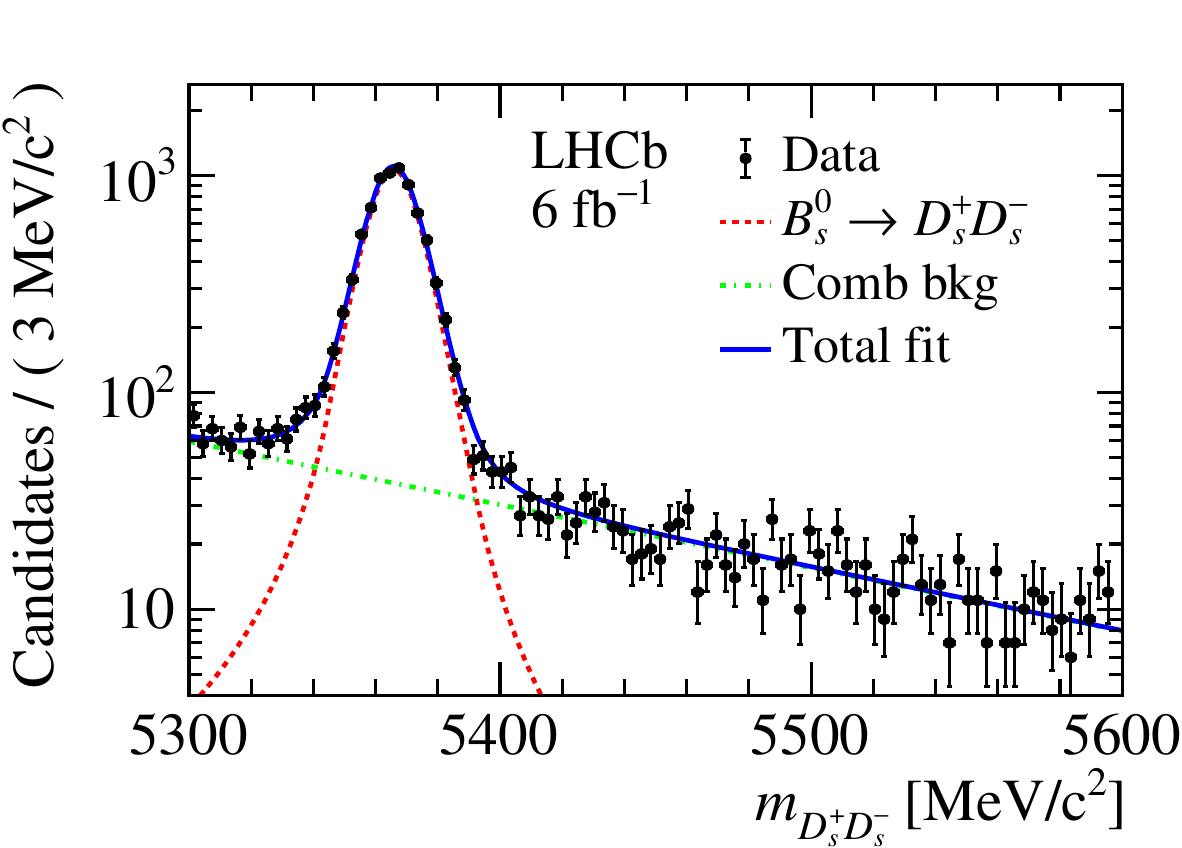}
        \includegraphics[width=0.49\linewidth]{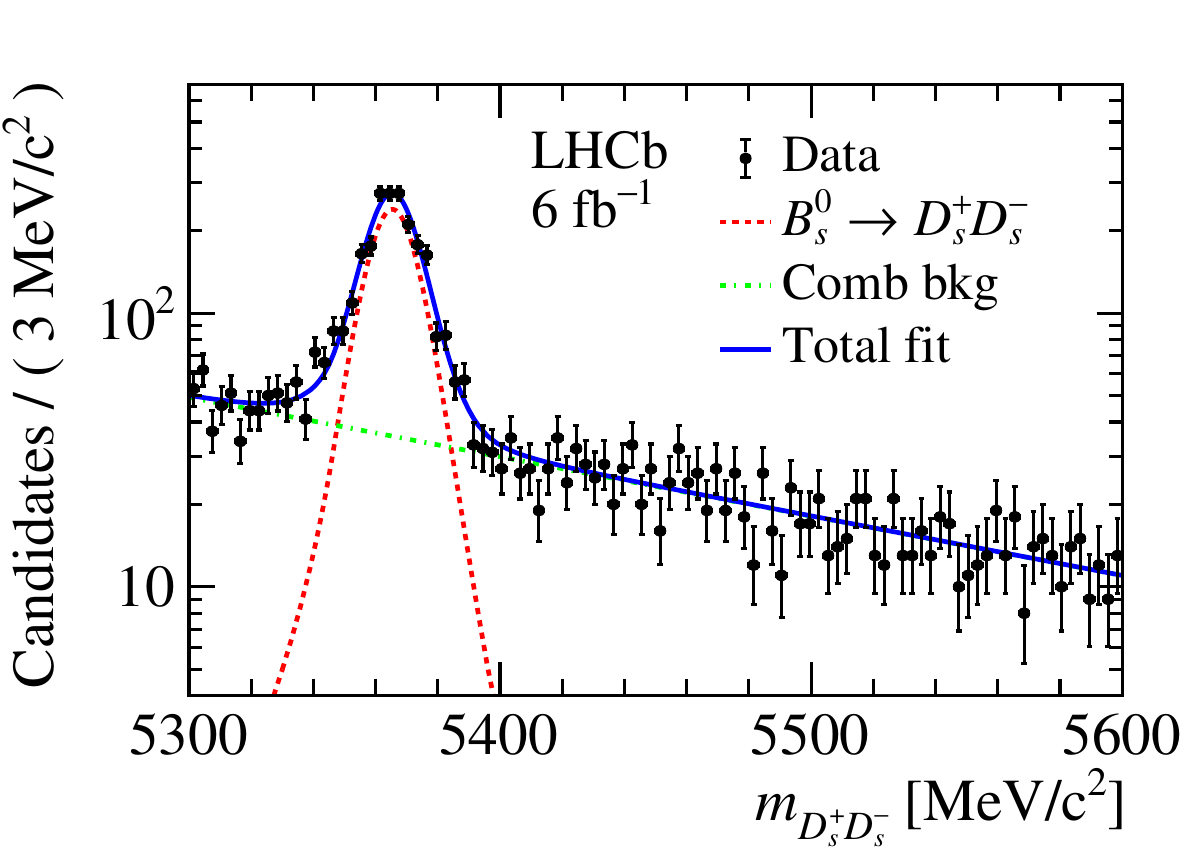}
        \includegraphics[width=0.49\linewidth]{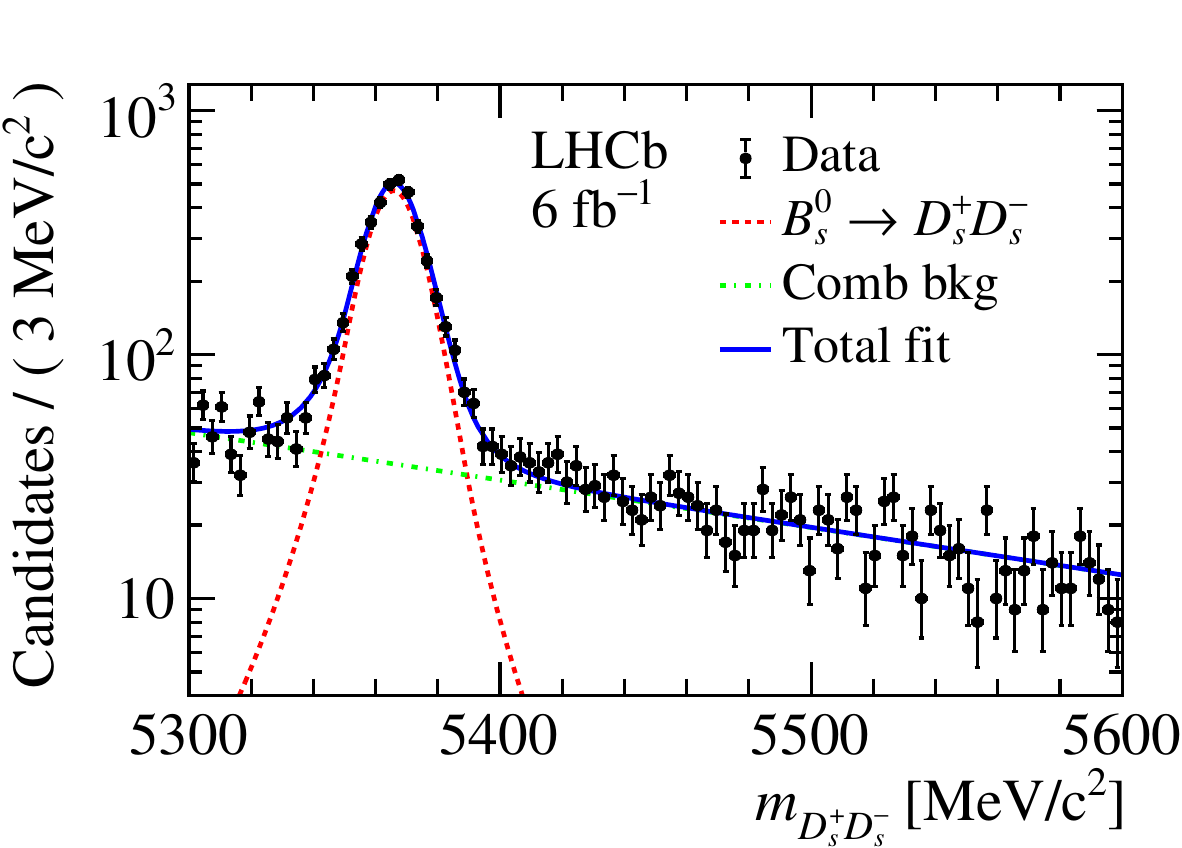}
        \vspace*{-0.5cm}
    \end{center}
    \caption{Invariant-mass distribution of $B_{s}^{0} \rightarrow D_{s}^{+} D_{s}^{-}$ decays.
        The data are shown as points and the full PDF is shown as a solid-blue line for
        (left) both $D_{s}^{\pm}$ candidates decaying through $D_{s}^{+} \rightarrow K^{-}K^{+}\pi^{+}$, (right) one $D_{s}^{\pm}$ candidate decaying through $D_{s}^{+} \rightarrow \pi^{-}K^{+}\pi^{+}$ and (bottom) one $D_{s}^{\pm}$ candidate decaying through $D_{s}^{+} \rightarrow \pi^{-}\pi^{+}\pi^{+}$.}
    \label{fig:massfits_bstodsds}
\end{figure}

% mass fits Bs2DsDs
The mass model in the \decay{\Bs}{\Dsp\Dsm} channel consists only of a signal component and
a combinatorial background component, which are parameterised as in the \mbox{\decay{\Bd}{\Dp\Dm}} fit.
Mass fits are performed separately for each final state.
Figures~\ref{fig:massfits_btodd} and \ref{fig:massfits_bstodsds} show the results of the fits to all  \decay{\Bd}{\Dp\Dm} and \decay{\Bs}{\Dsp\Dsm} final states, respectively.
The fits yield an overall number of $5\,695 \pm 100$ \decay{\Bd}{\Dp\Dm} and $13\,313 \pm 135$ \decay{\Bs}{\Dsp\Dsm} signal decays.

\section{Flavour tagging}
\label{sec:tagging}
For time-dependent \CP violation measurements of neutral \B mesons, the flavour of the meson at production is required.
At \lhcb the method used to determine the initial flavour is called flavour tagging.
These algorithms exploit the fact that in \proton\proton collisions, \bquark and \bquarkbar quarks are almost exclusively produced in pairs.
When the \bquark quark forms a \Bbar meson (and similarly the \bquarkbar quark forms a \B meson), additional particles are produced in the fragmentation process.
From the charges and types of these particles, the flavour of the signal \B meson at production can be inferred.
The tagging algorithm that uses charged pions or protons from the fragmentation process of the \bquark quark that leads to the \Bzb signal
is called the same-side (SS) tagger~\cite{LHCb-PAPER-2016-039}.
In the case of signal \Bsb mesons, charged kaons are used by the SS tagger~\cite{LHCb-PAPER-2015-056}.
The opposite-side (OS) tagger uses information from electrons and muons from semileptonic \bquark decays, kaons from
the \decay{\decay{\bquark}{\cquark}}{\squark} decay chain, secondary charm hadrons and the charges of tracks from the secondary vertex of the
other \bquark-hadron decay~\cite{LHCb-PAPER-2011-027,LHCb-PAPER-2015-027}.
Each algorithm $i$ provides individual tag decisions, $d_i$, and a predicted mistag, $\eta_i$, which is an estimate of the probability
that the tag decision is wrong.
The tag decision takes the values $-1$ for a \Bbar meson, $1$ for a \B meson and $0$ if no tag decision can be made.
The predicted mistag ranges from $0$ to $0.5$ and takes the value of $0.5$ for untagged events.
Each predicted mistag distribution is given by the output of a BDT that is trained on flavour-specific decays~\cite{Fazzini:2018dyq} and
has to be calibrated to represent the mistag probability, $\mistag_i(\eta_i)$, in the signal decay.
Flavour-specific control channels with kinematics similar to the signal are used to obtain a calibration curve.
This is found to be well-described by a linear function.
Following calibration, the individual taggers are combined separately for OS and SS cases, and the resulting mistag distributions are recalibrated.
These calibrations are used in the decay-time fit to determine the \CP-violation parameters to which the uncertainties on the calibration parameters are propagated through means of a Gaussian constraint.

To calibrate the SS and OS taggers of the \decay{\Bd}{\Dp\Dm} channel, as well as the OS tagger of the \decay{\Bs}{\Dsp\Dsm}
channel, \decay{\Bd}{\Dsp\Dm} decays are used.
These have very similar kinematics to the signal decays and the selection is very similar, as described in \cref{sec:selection}.
The SS kaon tagger used for \decay{\Bs}{\Dsp\Dsm} decays is calibrated with the
\decay{\Bs}{\Dsm\pip} channel.
A reweighting process is applied to ensure the calibration sample matches the distributions of the signal channel in the transverse momentum of the \Bs meson, the pseudorapidity, the number of tracks and the number of PVs.
Additionally, the compatibility of the calibration between \decay{\Bs}{\Dsm\pip} and \decay{\Bs}{\Dsp\Dsm} decays is verified by comparing the calibration parameters
determined using simulation.

The performance of the tagging algorithms is measured by the tagging power $\epsilon_\text{tag}D^2$, where $\epsilon_\text{tag}$
is the fraction of tagged candidates and $D=1-2\omega$ is the dilution factor introduced by the mistag probability, $\omega$.
The tagging power is a statistical dilution factor due to imperfect tagging, equivalent to an efficiency with respect to a sample with perfect tagging.
Overall tagging powers of $(6.28\pm0.11)\%$ in \decay{\Bd}{\Dp\Dm} and $(5.60\pm0.07)\%$ in \decay{\Bs}{\Dsp\Dsm} decays are achieved. 

\section{Decay-time fit}
\label{sec:decaytimefit}
An unbinned maximum-likelihood fit to the signal-weighted decay-time distribution is performed to determine the \CP-violation parameters.
In order to avoid experimenter bias, the values of the \CP-violation parameters were
not examined until the full procedure had been finalised.

The measured decay-time distribution of the \BdorBs candidates given the tag decisions $\vec{d} = (d_\text{OS}, d_\text{SS})$ and predicted mistags $\vec{\eta} = (\eta_\text{OS}, \eta_\text{SS})$ is described by the PDF
\begin{align}
    \label{eq:decaytimepdf}
    \mathcal{P}(t,\vec{d}\,|\,\vec{\eta}) = \epsilon(t) \cdot \left( \mathcal{B}(t',\vec{d}\,|\,\vec{\eta}) \otimes \mathcal{R}(t-t') \right) ,
\end{align}
where $\mathcal{B}(t',\vec{d}\,|\,\vec{\eta})$ describes the distribution of the true decay time $t'$, which is convolved with the decay-time resolution function
$\mathcal{R}(t-t')$, and the acceptance function $\epsilon(t)$ describes the total efficiency as a function of the reconstructed decay time.
The PDF describing the decay-time distribution can be deduced from \cref{eq:decayrate} and takes the general form
\begin{equation}
    \begin{aligned}
        \label{eq:decaytimepdf_phys}
        \mathcal{B}(t',\vec{d}\,|\,\vec{\eta}) \propto  e^{-t'/\tau}
         & \bigg( C^\text{eff}_{\cosh{}}(\vec{d}\,|\,\vec{\eta}) \cosh{\frac{\DG_{\quark} t'}{2}}
        + C^\text{eff}_{\sinh{}}(\vec{d}\,|\,\vec{\eta}) \sinh{\frac{\DG_{\quark} t'}{2}}         \\
         & - C^\text{eff}_{\cos{}}(\vec{d}\,|\,\vec{\eta}) \cos{\dm_{\quark} t'}
        + C^\text{eff}_{\sin{}}(\vec{d}\,|\,\vec{\eta}) \sin{\dm_{\quark} t'} \bigg).
    \end{aligned}
\end{equation}
The effective coefficients are given by
\begin{equation}
    \begin{aligned}
        C^\text{eff}_{\cosh{}} & = \phantom{D_f\bigg(}\Sigma(\vec{d}\,|\,\vec{\eta}) + A_\text{prod} \Delta(\vec{d}\,|\,\vec{\eta}), \quad
                               & C^\text{eff}_{\cos{}} = C_f \bigg(\Delta(\vec{d}\,|\,\vec{\eta}) + A_\text{prod} \Sigma(\vec{d}\,|\,\vec{\eta})\bigg),   \\
        C^\text{eff}_{\sinh{}} & = D_f \bigg(\Sigma(\vec{d}\,|\,\vec{\eta}) + A_\text{prod} \Delta(\vec{d}\,|\,\vec{\eta})\bigg),    \quad
                               & C^\text{eff}_{\sin{}}   = S_f \bigg(\Delta(\vec{d}\,|\,\vec{\eta}) + A_\text{prod} \Sigma(\vec{d}\,|\,\vec{\eta})\bigg),
    \end{aligned}
\end{equation}
where the production asymmetry $A_\text{prod} = (N_\BdorBsbar - N_\BdorBs)/(N_\BdorBsbar + N_\BdorBs)$ represents the difference
in the production rates of \BdorBsbar and \BdorBs mesons. The functions
\begin{equation}
\begin{aligned}
    \Sigma(\vec{d},\vec{\eta}) &= P(\vec{d},\vec{\eta}|\BdorBsbar) + P(\vec{d},\vec{\eta}|\BdorBs) \text{ and} \\
    \Delta(\vec{d},\vec{\eta}) &= P(\vec{d},\vec{\eta}|\BdorBsbar) - P(\vec{d},\vec{\eta}|\BdorBs)
\end{aligned}
\end{equation}
are dependent on the tagging calibration parameters, where $P(\vec{d},\vec{\eta}|\BdorBs)$ and $P(\vec{d},\vec{\eta}|\BdorBsbar)$ are the probabilities of observing the tagging decisions $\vec{d}$ and the predicted mistags $\vec{\eta}$, given the true flavour $\BdorBs$ or $\BdorBsbar$, respectively.

%%% b2dd
\subsection*{\decay{\Bd}{\Dp\Dm}}
The decay-time fit of \decay{\Bd}{\Dp\Dm} decays is insensitive to $C^\text{eff}_{\sinh{}}$ under the assumption that \DGd is zero.
% decay-time resolution
Moreover, due to the long oscillation period of the \Bd mesons, the decay-time resolution of around $52\fs$ has a very small impact on the \CP-violation parameters.
The decay-time resolution model consists of three Gaussian functions that have a common mean and different widths. The parameters of the model are determined from simulation and fixed in the fit to data.

% decay-time acceptance
The selection and reconstruction efficiency depends on the \Bd decay time due to displacement requirements made on the final-state particles and
a decrease in the reconstruction efficiency for tracks with large impact parameter with respect to the beamline~\cite{LHCb-PAPER-2013-065}.
The decay-time dependent efficiency is modeled by cubic-spline functions~\cite{karbach2014decay} with five knots at $(0.3, 0.5, 2.7, 6.3, 10.3)\ps$, whose positions were determined using simulation.
The spline coefficients are free to vary in the fit.

\begin{figure}[tb]
    \begin{center}
        \includegraphics[width=0.49\linewidth]{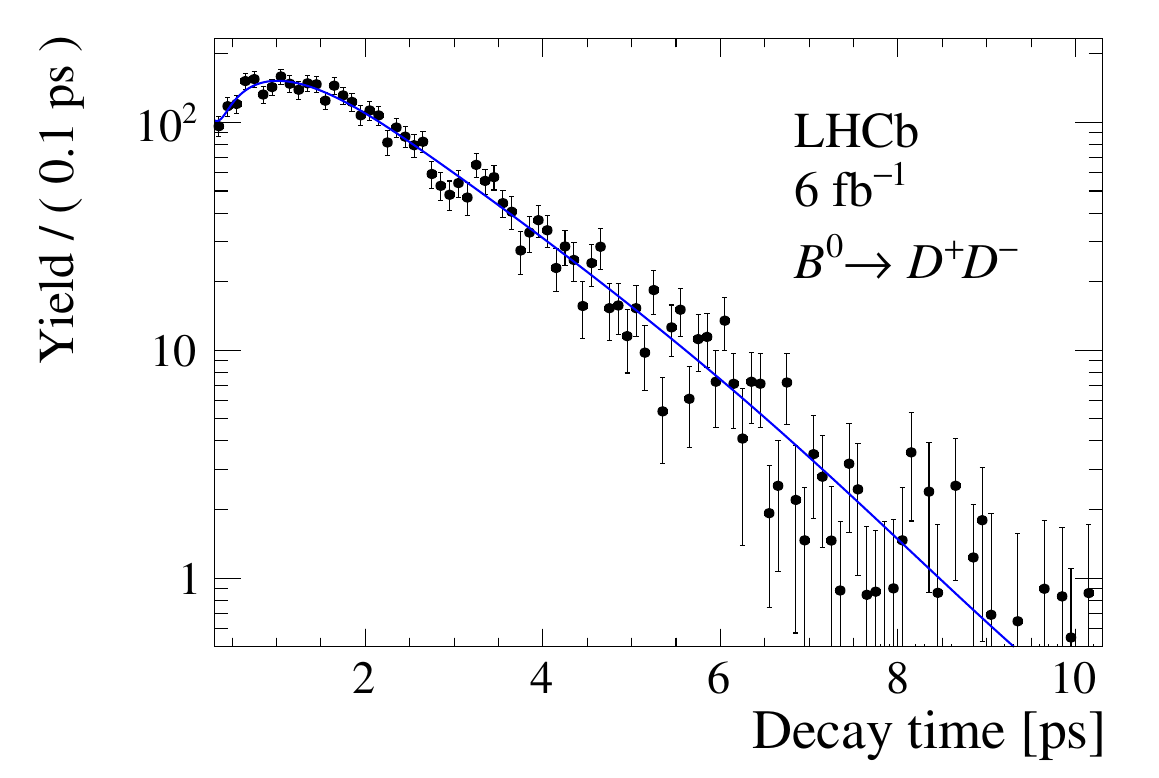}
        \includegraphics[width=0.49\linewidth]{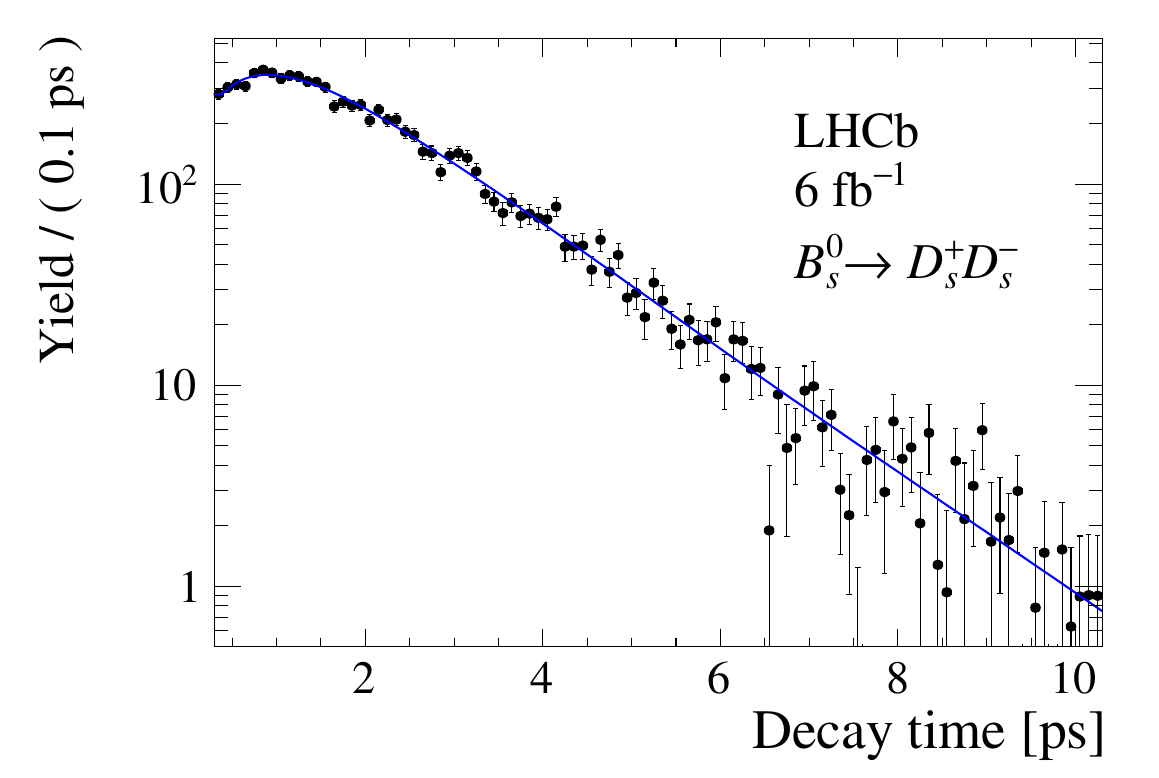}
    \end{center}
    \caption{Decay-time distribution of (left) $B^{0} \rightarrow D^{+} D^{-}$  and (right) $B_{s}^{0} \rightarrow D_{s}^{+} D_{s}^{-}$ candidates.
        The background-subtracted data are shown as points and the projection of the PDF is shown as a solid blue line.}
    \label{fig:decaytimefit}
\end{figure}

% input parameters
Gaussian constraints are used to account for the uncertainties on the tagging calibration parameters,
the \Bd lifetime, the oscillation frequency, \dmd, and the production asymmetry.
The world-average values are used for the external parameters~\cite{PDG2022}, while the production asymmetry is taken from
a similar time-dependent analysis of \decay{\Bd}{\Dstarpm\Dmp} decays~\cite{LHCb-PAPER-2019-036}.
The tagging efficiencies are free to vary in the decay-time fit.
%
% fit results
Figure~\ref{fig:decaytimefit} (left) shows the results of the decay-time fit for this channel.

%%% bs2dsds
\subsection*{\decay{\Bs}{\Dsp\Dsm}}
In the decay-time fit of \decay{\Bs}{\Dsp\Dsm} decays, the hyperbolic terms of \cref{eq:decaytimepdf_phys} can be measured provided that \DGs is not zero. Moreover,
the definitions from \cref{eq:cpparameters} are used to directly determine the parameters \phis and $|\lambda|$.
% decay-time acceptance and input parameters
The acceptance function, the tagging parameters and external parameters are treated in the same way as for the \decay{\Bd}{\Dp\Dm} decays.
In addition to the lifetime and the oscillation frequency, \dms, the decay-width difference \DGs is constrained in the fit to the world-average value~\cite{PDG2022}.
The value of the production asymmetry is taken from the control channel \decay{\Bs}{\Dsm\pip} as described in Ref.~\cite{LHCb-PAPER-2021-005}.

% decay-time resolution
Due to the high oscillation frequency of the \Bs meson, the decay-time resolution plays an important role. A per-event decay-time resolution is determined based on the per-event decay-time uncertainty estimated from the vertex fit, which is calibrated using a sample of \Dsm\pip candidates, with \decay{\Dsm}{\phiz(\Kp\Km)\pim}, and additional requirements imposed to suppress candidates produced in \B decays to negligible levels.
The measured decay time of the remaining candidates, which originate from the PV, is consistent with zero, and their distribution is used to assess  resolution and bias effects.
A linear fit to the measured and predicted decay-time resolution is performed.
A scale factor is then applied to translate the resulting calibration to the signal \decay{\Bs}{\Dsp\Dsm} mode. It is determined by comparing the decay-time resolution of \decay{\Bs}{\Dsm\pip} and \decay{\Bs}{\Dsp\Dsm} decays in simulation.
%
% fit results
Figure~\ref{fig:decaytimefit} (right) shows the results of the decay-time fit for this channel.

\begin{figure}[tb]
    \begin{center}
        \includegraphics[width=0.49\linewidth]{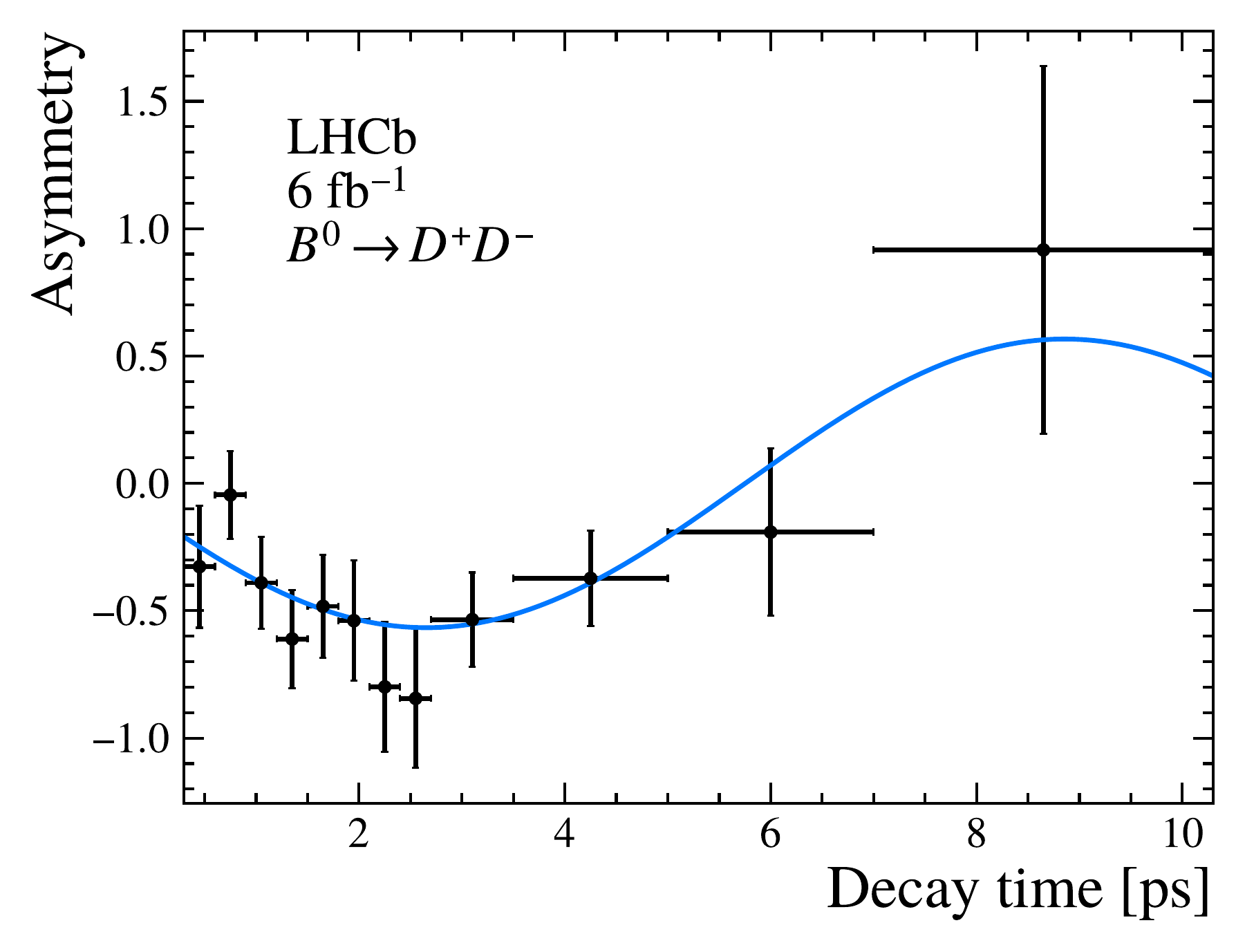}
        \includegraphics[width=0.49\linewidth]{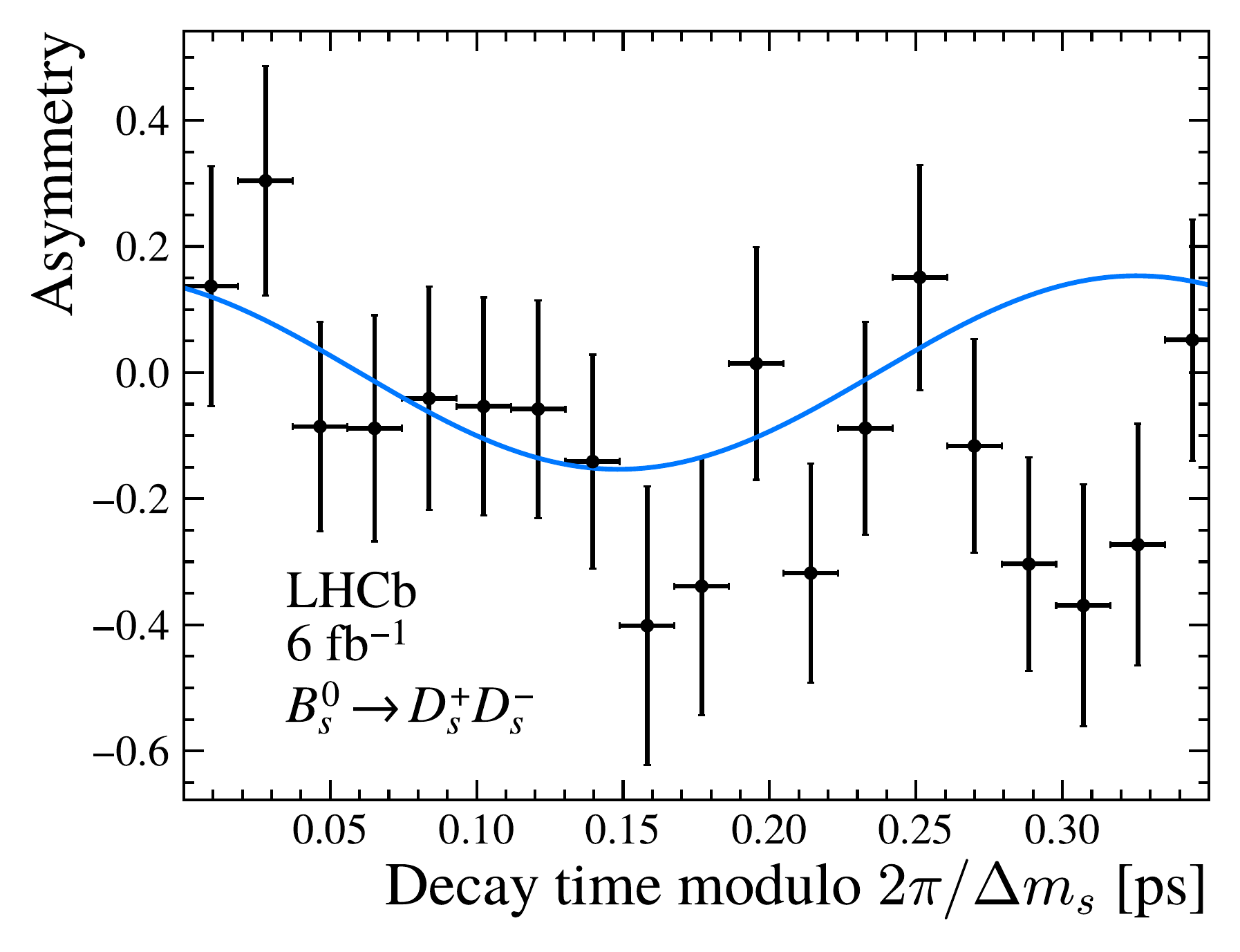}
    \end{center}
    \caption{Decay-time-dependent \CP asymmetry of (left) $B^{0} \rightarrow D^{+} D^{-}$ and (right) $B_{s}^{0} \rightarrow D_{s}^{+} D_{s}^{-}$ candidates.
        The asymmetry in the background-subtracted data is shown as points and the projection of the PDF is shown as a solid blue line.
        Due to the high oscillation frequency of the \Bs mesons, the corresponding distribution is folded onto one oscillation period.}
    \label{fig:asymmetry}
\end{figure}

The decay-time-dependent \CP asymmetry and the projection of the PDF are shown in \cref{fig:asymmetry} for (left) \decay{\Bd}{\Dp\Dm} and (right) \decay{\Bs}{\Dsp\Dsm} decays. The \CP asymmetry in each decay-time bin is given by $A^\CP = -(\sum_j w_jd_jD_j)/(\sum_j w_jD_j^2)$ with the tagging decision $d_j$, the tagging dilution $D_j$ and the signal weight $w_j$ obtained by the \sPlot method~\cite{LHCb-PAPER-2023-013}, for each candidate $j$.

\section{Systematic uncertainties and cross-checks}
\label{sec:systematics}
A variety of cross-checks are performed and potential sources of systematic uncertainties are considered.

% MC fit
The decay-time fit is performed on a simulated \decay{\Bd}{\Dp\Dm} sample using the same strategy for
the tagging calibration as for the fit to data.
A second fit is performed where instead of the reconstructed tagging, the truth information of the initial flavour
of the \Bd mesons is used.
Both results of the \CP-violation parameters agree with the generated values.

% splits
The decay-time fit is performed on several subsets of the data to test the consistency of the results.
The data subdivision is done according to the final state, magnet polarity, years of data taking and tagging information (OS only or SS only). Consistent results are found in all cases.

% bootstrapping
A bootstrapping procedure~\cite{efron:1979} is used to cross-check the statistical uncertainty from the decay-time fit to data.
A data set is created by randomly drawing candidates from the original sample until a certain number of candidates is reached that itself is drawn from a Poisson distribution with the expected number of candidates matching the original data sample.
This entails that the same candidate can be drawn multiple times.
The mass and decay-time fits are performed on this data set to first statistically subtract the background and then determine the \CP-violation parameters.
The residual of the fit result with respect to the baseline fit is stored and the whole procedure is repeated until the distribution of the residuals is not significantly affected by statistical fluctuations.
The statistical uncertainties from the fits to data are shown to be accurate as they are consistent with the standard deviations of the residuals, and the correlation coefficients lie within expectations.

% acceptance
A decay-time fit with a different set of knots for the acceptance function is performed.
The difference in the results with respect to the baseline fit is assigned as a systematic uncertainty.

% fit bias
To test the fit strategy, pseudoexperiments are performed.
In each pseudoexperiment, the mass and decay time are generated using the results of the baseline fit to data.
The background contributions are generated with a specific time dependence, assuming \CP symmetry for the \decay{\Bs}{\Dp\Dm} background.
Similar to the bootstrapping procedure, the baseline fitting procedure is performed on the pseudoexperiments and the residuals are collected.
For \decay{\Bd}{\Dp\Dm} decays, the mean values of the results are found to be consistent with the input values within the statistical uncertainties, while the fits to the \decay{\Bs}{\Dsp\Dsm} pseudoexperiments show a small bias of $-0.002$ in \phis and $0.008$ in $|\lambda_{\Dsp\Dsm}|$.
This is of the order of a few percent of the statistical uncertainty and is subtracted from the biases found in the following studies.

The following systematic uncertainties are determined using the same procedure, with the only difference being that an alternative model is used to generate pseudoexperiments in each case.
A bias in the distribution of the residuals is assigned as a systematic uncertainty.

% mass model
The sum of two Crystal Ball functions~\cite{Skwarnicki:1986xj}, with parameters obtained from a fit to simulation, is used in the pseudoexperiments to test the choice of the signal mass model.

% DGd
Since \DGd is fixed to zero in the decay-time fit of \decay{\Bd}{\Dp\Dm} decays, a systematic uncertainty is assigned for this assumption.
The value of \DGd is varied in the pseudoexperiments from the assumed value of zero by $\pm1\sigma$, where $\sigma$ is the uncertainty of the world average value of \DGd~\cite{PDG2024}.
The value of $D_{\Dp\Dm}$ is calculated from the normalisation condition $D_{\Dp\Dm} = \pm\sqrt{1 - S_{\Dp\Dm}^2 - C_{\Dp\Dm}^2}$ and the largest deviation is assigned as the systematic uncertainty.

% decay-time model 
In the \decay{\Bd}{\Dp\Dm} channel the decay-time-resolution model is determined on simulation.
Due to differences between simulation and data the resolution could be underestimated.
The effect of underestimating the resolution is tested by increasing the width of the resolution function by $10\%$ in the pseudoexperiments, which corresponds to the level measured in the \Bs system. It is found to be small and no further studies are considered.

In the \decay{\Bs}{\Dsp\Dsm} channel, \Dsm\pip candidates originating from the PV are used to determine a per-event resolution calibration.
Only \decay{\Dsm}{\phiz(\Kp\Km)\pim} decays are used and assumed to represent the resolution of the whole sample.
A second calibration is obtained using a sample of \decay{\Dsm}{\Kp\Km\pim} decays without specific requirements on the intermediate decays and used in the pseudoexperiments to assign a systematic uncertainty.

% decay-time bias
A decay-time bias caused by the misalignment of the vertex detector was observed in other \lhcb analyses of data taken during the same period~\cite{LHCb-PAPER-2021-005,LHCb-PAPER-2023-013} and confirmed in the present analysis.
Due to the low oscillation frequency of \Bd mesons, this has a negligible effect on the measurement of the \CP-violation parameters, as shown in Ref.~\cite{LHCb-PAPER-2021-005} and so is not evaluated here.
However, in \Bs decays, this bias could have a significant impact on the measurement.
To evaluate the effect, the mean of the resolution function in the generation of the pseudoexperiments is set to the largest observed bias.

%
% summary
The individual systematic uncertainties on the \CP-violation parameters are reported in \cref{tab:systematics} and summed in quadrature.
\begin{table}[tb]
    \centering
    \caption{Systematic uncertainties for the \decay{\Bd}{\Dp\Dm} and \decay{\Bs}{\Dsp\Dsm} channel. A dash (---) is used to denote that a systematic has not been evaluated. 
        The total systematic uncertainty is the quadratic sum of the individual uncertainties.}
    \label{tab:systematics}
    \begin{tabular}{lrrrr}
        \hline \noalign{\vskip 0.5mm}
        Source                & $S_{\Dp\Dm}$ & $C_{\Dp\Dm}$ & $\phis [\rad]$ & $|\lambda_{\Dsp\Dsm}|$ \\
        \noalign{\vskip 0.5mm}
        \hline
        Mass model            & $0.001$     & $0.005$      & $0.003$        & $0.005$                \\
        \DG                   & $0.010$     & $0.005$      & ---            & ---                    \\
        Decay-time resolution & $0.002$     & $0.007$      & $0.011$        & $0.027$                \\
        Decay-time bias       & ---         & ---          & $0.026$        & $0.014$                \\
        Acceptance function   & $0.001$     & $0.001$      & $<0.001$            & $0.001$                \\
        \hline
        Total                 & $0.010$     & $0.010$      & $0.028$        & $0.031$                \\
        \hline
    \end{tabular}
\end{table}
Compared to the previous \lhcb results~\cite{LHCb-Paper-2016-037,LHCb-PAPER-2014-051}, the systematic uncertainties in the \mbox{\decay{\Bd}{\Dp\Dm}} measurement are significantly reduced, while they are comparable for the \mbox{\decay{\Bs}{\Dsp\Dsm}} measurement.
The reduction in the first measurement is achieved by the improved suppression of backgrounds from single-charm, misidentified and partially reconstructed decays.
They are the leading sources of systematic uncertainties in the previous measurement and are reduced to negligible levels in this analysis.

\section{Results and interpretation}
\label{sec:results}
A flavour-tagged time-dependent analysis of \decay{\Bd}{\Dp\Dm} and \decay{\Bs}{\Dsp\Dsm} decays is performed
using proton-proton collision data collected by the \lhcb experiment during the years 2015 to 2018, corresponding to an integrated luminosity of 6\invfb.
Approximately 5\,700 \decay{\Bd}{\Dp\Dm} signal candidates are observed.
A fit to their decay-time distribution, including evaluation of systematic uncertainties, gives the final results
\begin{align}
    S_{\Dp\Dm} & = \SResultRunTwo, \nonumber           \\
    C_{\Dp\Dm} & = \phantom{-}\CResultRunTwo , \nonumber
\end{align}
with a statistical correlation between the two parameters of $\rho(S_{\Dp\Dm}, C_{\Dp\Dm}) = 0.472$.
The results and correlations of the external parameters from the decay-time fit are presented in \cref{app:ext_params}.
Wilks' theorem~\cite{Wilks:1938dza} is used to determine the significance of the result, excluding systematic uncertainties.
The hypothesis of \CP symmetry, corresponding to $S_{\Dp\Dm} = C_{\Dp\Dm} = 0$, can be rejected by more than six standard deviations.
The values are consistent with previous results from \lhcb and \babar~\cite{PhysRevD.79.032002}, which
correspond to a small contribution from higher-order SM corrections.
Thus, this measurement will move the world average further away from the \belle measurement,
which lies outside the physical region~\cite{PhysRevD.85.091106}.

The result is combined with the previous \lhcb measurement in this channel~\cite{LHCb-Paper-2016-037} using the \texttt{GammaCombo} package~\cite{GammaCombo,*LHCb-PAPER-2016-032}.
Due to the small effect of the external parameters on the result, the two measurements are
assumed to be uncorrelated and the combined values are
\begin{align}
    S_{\Dp\Dm} & = \SResultCombined, \nonumber         \\
    C_{\Dp\Dm} & = \phantom{-}\CResultCombined,\nonumber
\end{align}
with a statistical correlation between the two parameters of $\rho(S_{\Dp\Dm}, C_{\Dp\Dm}) = 0.474$.

Approximately 13\,000 \decay{\Bs}{\Dsp\Dsm} signal candidates are observed and the final results of the
decay-time fit and the systematic uncertainties are
\begin{align}
    \phis                & = \phisResultRunTwo, \nonumber              \\
    |\lambda_{\Dsp\Dsm}| & = \phantom{-}\lambdaResultRunTwo , \nonumber
\end{align}
with a statistical correlation between the two parameters of $\rho(\phis, |\lambda_{\Dsp\Dsm}|) = -0.007$.
Further information on the results of the decay-time fit is shown in \cref{app:ext_params}.
This result is consistent with, and more precise than, the previous \lhcb measurement~\cite{LHCb-PAPER-2014-051}.
The combination with the previous \lhcb measurement, following the same strategy as for the \decay{\Bd}{\Dp\Dm} decays, yields the values
\begin{align}
    \phis                & = \phisResultCombined, \nonumber     \\
    |\lambda_{\Dsp\Dsm}| & = \phantom{-} \lambdaResultCombined ,\nonumber
\end{align}
with a statistical correlation between the two parameters of $\rho(\phis, |\lambda_{\Dsp\Dsm}|) = 0.005$.
The values are consistent with \CP symmetry in the \decay{\Bs}{\Dsp\Dsm} channel.

These results can be used in combination with other \decay{\B}{\D\D} measurements to perform a global analysis
and extract SM parameters as has previously been performed in Ref.~\cite{Davies2024}.
They represent the most precise single measurements of the \CP-violation parameters in their respective channels and the combined results supersede the previous \lhcb measurements.
For the first time, \CP symmetry can be excluded by more than six standard deviations in a single measurement of \decay{\Bd}{\Dp\Dm} decays.

% Do not include this in any draft (just for information in the template)
%\input{acknowledgements_intro}
% Comment this in for paper drafts; do not include this in analysis note, conference and figure reports
\section*{Acknowledgements}
%
% These Acknowledgements valid from 3-May-2019
%
\noindent We express our gratitude to our colleagues in the CERN
accelerator departments for the excellent performance of the LHC. We
thank the technical and administrative staff at the LHCb
institutes.
We acknowledge support from CERN and from the national agencies:
CAPES, CNPq, FAPERJ and FINEP (Brazil); 
MOST and NSFC (China); 
CNRS/IN2P3 (France); 
BMBF, DFG and MPG (Germany); 
INFN (Italy); 
NWO (Netherlands); 
MNiSW and NCN (Poland); 
MCID/IFA (Romania); 
%MSHE (Russia); 
MICIU and AEI (Spain);
SNSF and SER (Switzerland); 
NASU (Ukraine); 
STFC (United Kingdom); 
DOE NP and NSF (USA).
We acknowledge the computing resources that are provided by CERN, IN2P3
(France), KIT and DESY (Germany), INFN (Italy), SURF (Netherlands),
PIC (Spain), GridPP (United Kingdom), 
%RRCKI and Yandex LLC (Russia), 
CSCS (Switzerland), IFIN-HH (Romania), CBPF (Brazil),
and Polish WLCG (Poland).
We are indebted to the communities behind the multiple open-source
software packages on which we depend.
Individual groups or members have received support from
ARC and ARDC (Australia);
Key Research Program of Frontier Sciences of CAS, CAS PIFI, CAS CCEPP, 
Fundamental Research Funds for the Central Universities, 
and Sci. \& Tech. Program of Guangzhou (China);
Minciencias (Colombia);
EPLANET, Marie Sk\l{}odowska-Curie Actions, ERC and NextGenerationEU (European Union);
A*MIDEX, ANR, IPhU and Labex P2IO, and R\'{e}gion Auvergne-Rh\^{o}ne-Alpes (France);
%RFBR, RSF and Yandex LLC (Russia);
AvH Foundation (Germany);
ICSC (Italy); 
%GVA, XuntaGal, GENCAT, Inditex, InTalent and Prog.~Atracci\'on Talento, CM (Spain);
Severo Ochoa and Mar\'ia de Maeztu Units of Excellence, GVA, XuntaGal, GENCAT, InTalent-Inditex and Prog. ~Atracci\'on Talento CM (Spain);
SRC (Sweden);
the Leverhulme Trust, the Royal Society
 and UKRI (United Kingdom).

\clearpage
\section*{Appendices}

\appendix
\section{Results and correlations of external parameters}
\label{app:ext_params}
\begin{table}[!htb]
    \centering
    \caption{Results of the external parameters from the decay-time fit to \decay{\Bd}{\Dp\Dm} data.}
    \label{tab:ext_results_btodd}
    \begin{tabular}{lcc}
        \hline
        Parameter     & Input value       & Fit result \\
        \hline
        \dmd [\invps] & $0.5065\pm0.0019$ & $0.5065\pm0.0019$ \\
        \tauBd [\ps]  & $1.519\pm0.004$   & $1.519\pm0.004$  \\
        \hline
    \end{tabular}
\end{table}
\begin{table}[!htb]
    \centering
    \caption{Correlation matrix of the \CP parameters and the external parameters from the decay-time fit to \decay{\Bd}{\Dp\Dm} data.}
    \label{tab:correlations_btodd}
    \begin{tabular}{lrrrr}
        \hline
        {}           & $S_{\Dp\Dm}$ & $C_{\Dp\Dm}$ & \dmd     & \tauBd   \\
        \hline
        $S_{\Dp\Dm}$ & $1.000$      & $0.472$      & $-0.014$ & $<0.001$ \\
        $C_{\Dp\Dm}$ &              & $1.000$      & $-0.022$ & $<0.001$ \\
        \dmd         &              &              & $1.000$  & $<0.001$ \\
        \tauBd       &              &              &          & $1.000$  \\
        \hline
    \end{tabular}
\end{table}
\begin{table}[!htb]
    \centering
    \caption{Results of the external parameters from the decay-time fit to \decay{\Bs}{\Dsp\Dsm} data.}
    \label{tab:ext_results_bstodsds}
    \begin{tabular}{lr@{\:$\pm$\:}lr@{\:$\pm$\:}l}
        \hline
        Parameter         & \multicolumn{2}{c}{Input value} & \multicolumn{2}{c}{Fit result}  \\
        \hline
        \DGs    [\invps]  & $0.083$ & $0.005$  & $0.083$ & $0.005$ \\
        \dms     [\invps] & $17.765$ & $0.006$ & $17.765$ & $0.006$ \\
        \tauBs    [\ps]   & $1.521$ & $0.005$ & $1.521$ & $0.005$ \\
        \hline
    \end{tabular}
\end{table}
\begin{table}[!htb]
    \centering
    \caption{Correlation matrix of the \CP parameters and the external parameters from the decay-time fit to \decay{\Bs}{\Dsp\Dsm}.}
    \label{tab:correlations_bstodsds}
    \begin{tabular}{lrrrrr}
        \hline \noalign{\vskip 0.5mm}
        {}          & \phis    & $\lambda_{\Dsp\Dsm}$ & \dms     & \DGs     & \tauBs   \\
        \noalign{\vskip 0.5mm}
        \hline
        \phis                &   $1.000$ &    $-0.007$ &  $-0.010$ &   $0.001$ &  $<0.001$ \\
        $\lambda_{\Dsp\Dsm}$ &           &     $1.000$ &  $-0.018$ &  $-0.010$ &  $<0.001$ \\
        \dms                 &           &             &   $1.000$ &  $<0.001$ &  $<0.001$ \\
        \DGs                 &           &             &           &   $1.000$ &  $<0.001$ \\
        \tauBs               &           &             &           &           &   $1.000$ \\
        \hline
    \end{tabular}
\end{table}
\clearpage

% This should be taken out in the final paper
%\input{supplementary-app}

\addcontentsline{toc}{section}{References}
%\setboolean{inbibliography}{true}
\bibliographystyle{LHCb}
\bibliography{main,standard,LHCb-PAPER,LHCb-CONF,LHCb-DP,LHCb-TDR}

\ifx\mcitethebibliography\mciteundefinedmacro
\PackageError{LHCb.bst}{mciteplus.sty has not been loaded}
{This bibstyle requires the use of the mciteplus package.}\fi
\providecommand{\href}[2]{#2}
\begin{mcitethebibliography}{10}
\mciteSetBstSublistMode{n}
\mciteSetBstMaxWidthForm{subitem}{\alph{mcitesubitemcount})}
\mciteSetBstSublistLabelBeginEnd{\mcitemaxwidthsubitemform\space}
{\relax}{\relax}

\bibitem{Fleischer_2007}
R.~Fleischer, \ifthenelse{\boolean{articletitles}}{\emph{{Exploring CP
  violation and penguin effects through $B^0_d$→\Dp\Dm and
  \mbox{\Bs→\Dsp\Dsm}}},
  }{}\href{https://doi.org/10.1140/epjc/s10052-007-0341-4}{Eur.\ Phys.\ J.\
  \textbf{C51} (2007) 849},
  \href{http://arxiv.org/abs/0705.4421}{{\normalfont\ttfamily
  arXiv:0705.4421}}\relax
\mciteBstWouldAddEndPuncttrue
\mciteSetBstMidEndSepPunct{\mcitedefaultmidpunct}
{\mcitedefaultendpunct}{\mcitedefaultseppunct}\relax
\EndOfBibitem
\bibitem{Fleischer_1999}
R.~Fleischer, \ifthenelse{\boolean{articletitles}}{\emph{{Extracting $\gamma$
  from ${{B_{s(d)} }}\to {J}/\psi{{K_S}}$ and ${{B_{d(s)}}}\to{{D^{+}_{d(s)}}}
  {{D^{-}_{d(s)}}}$}}, }{}\href{https://doi.org/10.1007/s100529900099}{Eur.\
  Phys.\ J.\  \textbf{C10} (1999) 299},
  \href{http://arxiv.org/abs/hep-ph/9903455}{{\normalfont\ttfamily
  arXiv:hep-ph/9903455}}\relax
\mciteBstWouldAddEndPuncttrue
\mciteSetBstMidEndSepPunct{\mcitedefaultmidpunct}
{\mcitedefaultendpunct}{\mcitedefaultseppunct}\relax
\EndOfBibitem
\bibitem{Jung_2015}
M.~Jung and S.~Schacht, \ifthenelse{\boolean{articletitles}}{\emph{{Standard
  model predictions and new physics sensitivity in
  $B\ensuremath{\rightarrow}DD$ decays}},
  }{}\href{https://doi.org/10.1103/PhysRevD.91.034027}{Phys.\ Rev.\
  \textbf{D91} (2015) 034027},
  \href{http://arxiv.org/abs/1410.8396}{{\normalfont\ttfamily
  arXiv:1410.8396}}\relax
\mciteBstWouldAddEndPuncttrue
\mciteSetBstMidEndSepPunct{\mcitedefaultmidpunct}
{\mcitedefaultendpunct}{\mcitedefaultseppunct}\relax
\EndOfBibitem
\bibitem{Bel_2015}
L.~Bel {\em et~al.}, \ifthenelse{\boolean{articletitles}}{\emph{{Anatomy of
  \B→\D\Db decays}}, }{}\href{https://doi.org/10.1007/jhep07(2015)108}{JHEP
  \textbf{7} (2015) 108},
  \href{http://arxiv.org/abs/1505.01361}{{\normalfont\ttfamily
  arXiv:1505.01361}}\relax
\mciteBstWouldAddEndPuncttrue
\mciteSetBstMidEndSepPunct{\mcitedefaultmidpunct}
{\mcitedefaultendpunct}{\mcitedefaultseppunct}\relax
\EndOfBibitem
\bibitem{Davies2024}
J.~Davies, M.~Jung, and S.~Schacht,
  \ifthenelse{\boolean{articletitles}}{\emph{{$\Bbar\to\Db\D$ decays and the
  extraction of $f_d/f_u$ at hadron colliders}},
  }{}\href{https://doi.org/10.1007/JHEP01(2024)191}{JHEP \textbf{2024} (2024)
  191}, \href{http://arxiv.org/abs/2311.16952}{{\normalfont\ttfamily
  arXiv:2311.16952}}\relax
\mciteBstWouldAddEndPuncttrue
\mciteSetBstMidEndSepPunct{\mcitedefaultmidpunct}
{\mcitedefaultendpunct}{\mcitedefaultseppunct}\relax
\EndOfBibitem
\bibitem{Cabibbo:1963yz}
N.~Cabibbo, \ifthenelse{\boolean{articletitles}}{\emph{{Unitary symmetry and
  leptonic decays}},
  }{}\href{https://doi.org/10.1103/PhysRevLett.10.531}{Phys.\ Rev.\ Lett.\
  \textbf{10} (1963) 531}\relax
\mciteBstWouldAddEndPuncttrue
\mciteSetBstMidEndSepPunct{\mcitedefaultmidpunct}
{\mcitedefaultendpunct}{\mcitedefaultseppunct}\relax
\EndOfBibitem
\bibitem{Kobayashi:1973fv}
M.~Kobayashi and T.~Maskawa,
  \ifthenelse{\boolean{articletitles}}{\emph{{\CP-violation in the
  renormalizable theory of weak interaction}},
  }{}\href{https://doi.org/10.1143/PTP.49.652}{Prog.\ Theor.\ Phys.\
  \textbf{49} (1973) 652}\relax
\mciteBstWouldAddEndPuncttrue
\mciteSetBstMidEndSepPunct{\mcitedefaultmidpunct}
{\mcitedefaultendpunct}{\mcitedefaultseppunct}\relax
\EndOfBibitem
\bibitem{LHCb-PAPER-2023-013}
LHCb collaboration, R.~Aaij {\em et~al.},
  \ifthenelse{\boolean{articletitles}}{\emph{{Measurement of $\CP$ violation in
  \mbox{$\Bz \rightarrow \psires (\to \ellell) \KS(\to \pip \pim)$} decays}},
  }{}\href{https://doi.org/10.1103/PhysRevLett.132.021801}{Phys.\ Rev.\ Lett.\
  \textbf{132} (2024) 021801},
  \href{http://arxiv.org/abs/2309.09728}{{\normalfont\ttfamily
  arXiv:2309.09728}}\relax
\mciteBstWouldAddEndPuncttrue
\mciteSetBstMidEndSepPunct{\mcitedefaultmidpunct}
{\mcitedefaultendpunct}{\mcitedefaultseppunct}\relax
\EndOfBibitem
\bibitem{LHCb-Paper-2016-037}
LHCb collaboration, R.~Aaij {\em et~al.},
  \ifthenelse{\boolean{articletitles}}{\emph{{Measurement of \CP violation in
  \mbox{\decay{\B}{\Dp\Dm}} decays}},
  }{}\href{https://doi.org/10.1103/PhysRevLett.117.261801}{Phys.\ Rev.\ Lett.\
  \textbf{117} (2016) 261801},
  \href{http://arxiv.org/abs/1608.06620}{{\normalfont\ttfamily
  arXiv:1608.06620}}\relax
\mciteBstWouldAddEndPuncttrue
\mciteSetBstMidEndSepPunct{\mcitedefaultmidpunct}
{\mcitedefaultendpunct}{\mcitedefaultseppunct}\relax
\EndOfBibitem
\bibitem{LHCb-PAPER-2014-051}
LHCb collaboration, R.~Aaij {\em et~al.},
  \ifthenelse{\boolean{articletitles}}{\emph{{Measurement of the \CP-violating
  phase \phis in \mbox{\decay{\Bsb}{\Dsp\Dsm}} decays}},
  }{}\href{https://doi.org/10.1103/PhysRevLett.113.211801}{Phys.\ Rev.\ Lett.\
  \textbf{113} (2014) 211801},
  \href{http://arxiv.org/abs/1409.4619}{{\normalfont\ttfamily
  arXiv:1409.4619}}\relax
\mciteBstWouldAddEndPuncttrue
\mciteSetBstMidEndSepPunct{\mcitedefaultmidpunct}
{\mcitedefaultendpunct}{\mcitedefaultseppunct}\relax
\EndOfBibitem
\bibitem{PhysRevD.79.032002}
\babar collaboration, B.~Aubert {\em et~al.},
  \ifthenelse{\boolean{articletitles}}{\emph{{Measurements of time-dependent
  \CP asymmetries in
  ${B}^{0}\ensuremath{\rightarrow}{D}^{(*)+}{D}^{(*)\ensuremath{-}}$ decays}},
  }{}\href{https://doi.org/10.1103/PhysRevD.79.032002}{Phys.\ Rev.\
  \textbf{D79} (2009) 032002},
  \href{http://arxiv.org/abs/0808.1866}{{\normalfont\ttfamily
  arXiv:0808.1866}}\relax
\mciteBstWouldAddEndPuncttrue
\mciteSetBstMidEndSepPunct{\mcitedefaultmidpunct}
{\mcitedefaultendpunct}{\mcitedefaultseppunct}\relax
\EndOfBibitem
\bibitem{PhysRevD.85.091106}
Belle collaboration, M.~R\"ohrken {\em et~al.},
  \ifthenelse{\boolean{articletitles}}{\emph{{Measurements of branching
  fractions and time-dependent \CP violating asymmetries in
  ${B}^{0}\ensuremath{\rightarrow}{D}^{(*)\ifmmode\pm\else\textpm\fi{}}{D}^{\ensuremath{\mp}}$
  decays}}, }{}\href{https://doi.org/10.1103/PhysRevD.85.091106}{Phys.\ Rev.\
  \textbf{D85} (2012) 091106},
  \href{http://arxiv.org/abs/1203.6647}{{\normalfont\ttfamily
  arXiv:1203.6647}}\relax
\mciteBstWouldAddEndPuncttrue
\mciteSetBstMidEndSepPunct{\mcitedefaultmidpunct}
{\mcitedefaultendpunct}{\mcitedefaultseppunct}\relax
\EndOfBibitem
\bibitem{LHCb-DP-2008-001}
LHCb collaboration, A.~A. Alves~Jr.\ {\em et~al.},
  \ifthenelse{\boolean{articletitles}}{\emph{{The \lhcb detector at the LHC}},
  }{}\href{https://doi.org/10.1088/1748-0221/3/08/S08005}{JINST \textbf{3}
  (2008) S08005}\relax
\mciteBstWouldAddEndPuncttrue
\mciteSetBstMidEndSepPunct{\mcitedefaultmidpunct}
{\mcitedefaultendpunct}{\mcitedefaultseppunct}\relax
\EndOfBibitem
\bibitem{LHCb-DP-2014-002}
LHCb collaboration, R.~Aaij {\em et~al.},
  \ifthenelse{\boolean{articletitles}}{\emph{{LHCb detector performance}},
  }{}\href{https://doi.org/10.1142/S0217751X15300227}{Int.\ J.\ Mod.\ Phys.\
  \textbf{A30} (2015) 1530022},
  \href{http://arxiv.org/abs/1412.6352}{{\normalfont\ttfamily
  arXiv:1412.6352}}\relax
\mciteBstWouldAddEndPuncttrue
\mciteSetBstMidEndSepPunct{\mcitedefaultmidpunct}
{\mcitedefaultendpunct}{\mcitedefaultseppunct}\relax
\EndOfBibitem
\bibitem{LHCb-DP-2014-001}
R.~Aaij {\em et~al.}, \ifthenelse{\boolean{articletitles}}{\emph{{Performance
  of the LHCb Vertex Locator}},
  }{}\href{https://doi.org/10.1088/1748-0221/9/09/P09007}{JINST \textbf{9}
  (2014) P09007}, \href{http://arxiv.org/abs/1405.7808}{{\normalfont\ttfamily
  arXiv:1405.7808}}\relax
\mciteBstWouldAddEndPuncttrue
\mciteSetBstMidEndSepPunct{\mcitedefaultmidpunct}
{\mcitedefaultendpunct}{\mcitedefaultseppunct}\relax
\EndOfBibitem
\bibitem{LHCb-DP-2017-001}
P.~d'Argent {\em et~al.}, \ifthenelse{\boolean{articletitles}}{\emph{{Improved
  performance of the LHCb Outer Tracker in LHC Run 2}},
  }{}\href{https://doi.org/10.1088/1748-0221/12/11/P11016}{JINST \textbf{12}
  (2017) P11016}, \href{http://arxiv.org/abs/1708.00819}{{\normalfont\ttfamily
  arXiv:1708.00819}}\relax
\mciteBstWouldAddEndPuncttrue
\mciteSetBstMidEndSepPunct{\mcitedefaultmidpunct}
{\mcitedefaultendpunct}{\mcitedefaultseppunct}\relax
\EndOfBibitem
\bibitem{LHCb-DP-2012-003}
M.~Adinolfi {\em et~al.},
  \ifthenelse{\boolean{articletitles}}{\emph{{Performance of the \lhcb RICH
  detector at the LHC}},
  }{}\href{https://doi.org/10.1140/epjc/s10052-013-2431-9}{Eur.\ Phys.\ J.\
  \textbf{C73} (2013) 2431},
  \href{http://arxiv.org/abs/1211.6759}{{\normalfont\ttfamily
  arXiv:1211.6759}}\relax
\mciteBstWouldAddEndPuncttrue
\mciteSetBstMidEndSepPunct{\mcitedefaultmidpunct}
{\mcitedefaultendpunct}{\mcitedefaultseppunct}\relax
\EndOfBibitem
\bibitem{LHCb-DP-2012-002}
A.~A. Alves~Jr.\ {\em et~al.},
  \ifthenelse{\boolean{articletitles}}{\emph{{Performance of the LHCb muon
  system}}, }{}\href{https://doi.org/10.1088/1748-0221/8/02/P02022}{JINST
  \textbf{8} (2013) P02022},
  \href{http://arxiv.org/abs/1211.1346}{{\normalfont\ttfamily
  arXiv:1211.1346}}\relax
\mciteBstWouldAddEndPuncttrue
\mciteSetBstMidEndSepPunct{\mcitedefaultmidpunct}
{\mcitedefaultendpunct}{\mcitedefaultseppunct}\relax
\EndOfBibitem
\bibitem{Sjostrand:2007gs}
T.~Sj\"{o}strand, S.~Mrenna, and P.~Skands,
  \ifthenelse{\boolean{articletitles}}{\emph{{A brief introduction to PYTHIA
  8.1}}, }{}\href{https://doi.org/10.1016/j.cpc.2008.01.036}{Comput.\ Phys.\
  Commun.\  \textbf{178} (2008) 852},
  \href{http://arxiv.org/abs/0710.3820}{{\normalfont\ttfamily
  arXiv:0710.3820}}\relax
\mciteBstWouldAddEndPuncttrue
\mciteSetBstMidEndSepPunct{\mcitedefaultmidpunct}
{\mcitedefaultendpunct}{\mcitedefaultseppunct}\relax
\EndOfBibitem
\bibitem{Sjostrand:2006za}
T.~Sj\"{o}strand, S.~Mrenna, and P.~Skands,
  \ifthenelse{\boolean{articletitles}}{\emph{{PYTHIA 6.4 physics and manual}},
  }{}\href{https://doi.org/10.1088/1126-6708/2006/05/026}{JHEP \textbf{05}
  (2006) 026}, \href{http://arxiv.org/abs/hep-ph/0603175}{{\normalfont\ttfamily
  arXiv:hep-ph/0603175}}\relax
\mciteBstWouldAddEndPuncttrue
\mciteSetBstMidEndSepPunct{\mcitedefaultmidpunct}
{\mcitedefaultendpunct}{\mcitedefaultseppunct}\relax
\EndOfBibitem
\bibitem{LHCb-PROC-2010-056}
I.~Belyaev {\em et~al.}, \ifthenelse{\boolean{articletitles}}{\emph{{Handling
  of the generation of primary events in Gauss, the LHCb simulation
  framework}}, }{}\href{https://doi.org/10.1088/1742-6596/331/3/032047}{J.\
  Phys.\ Conf.\ Ser.\  \textbf{331} (2011) 032047}\relax
\mciteBstWouldAddEndPuncttrue
\mciteSetBstMidEndSepPunct{\mcitedefaultmidpunct}
{\mcitedefaultendpunct}{\mcitedefaultseppunct}\relax
\EndOfBibitem
\bibitem{Lange:2001uf}
D.~J. Lange, \ifthenelse{\boolean{articletitles}}{\emph{{The EvtGen particle
  decay simulation package}},
  }{}\href{https://doi.org/10.1016/S0168-9002(01)00089-4}{Nucl.\ Instrum.\
  Meth.\  \textbf{A462} (2001) 152}\relax
\mciteBstWouldAddEndPuncttrue
\mciteSetBstMidEndSepPunct{\mcitedefaultmidpunct}
{\mcitedefaultendpunct}{\mcitedefaultseppunct}\relax
\EndOfBibitem
\bibitem{davidson2015photos}
N.~Davidson, T.~Przedzinski, and Z.~Was,
  \ifthenelse{\boolean{articletitles}}{\emph{{PHOTOS interface in C++:
  Technical and physics documentation}},
  }{}\href{https://doi.org/https://doi.org/10.1016/j.cpc.2015.09.013}{Comp.\
  Phys.\ Comm.\  \textbf{199} (2016) 86},
  \href{http://arxiv.org/abs/1011.0937}{{\normalfont\ttfamily
  arXiv:1011.0937}}\relax
\mciteBstWouldAddEndPuncttrue
\mciteSetBstMidEndSepPunct{\mcitedefaultmidpunct}
{\mcitedefaultendpunct}{\mcitedefaultseppunct}\relax
\EndOfBibitem
\bibitem{Allison:2006ve}
Geant4 collaboration, J.~Allison {\em et~al.},
  \ifthenelse{\boolean{articletitles}}{\emph{{Geant4 developments and
  applications}}, }{}\href{https://doi.org/10.1109/TNS.2006.869826}{IEEE
  Trans.\ Nucl.\ Sci.\  \textbf{53} (2006) 270}\relax
\mciteBstWouldAddEndPuncttrue
\mciteSetBstMidEndSepPunct{\mcitedefaultmidpunct}
{\mcitedefaultendpunct}{\mcitedefaultseppunct}\relax
\EndOfBibitem
\bibitem{Agostinelli:2002hh}
Geant4 collaboration, S.~Agostinelli {\em et~al.},
  \ifthenelse{\boolean{articletitles}}{\emph{{Geant4: A simulation toolkit}},
  }{}\href{https://doi.org/10.1016/S0168-9002(03)01368-8}{Nucl.\ Instrum.\
  Meth.\  \textbf{A506} (2003) 250}\relax
\mciteBstWouldAddEndPuncttrue
\mciteSetBstMidEndSepPunct{\mcitedefaultmidpunct}
{\mcitedefaultendpunct}{\mcitedefaultseppunct}\relax
\EndOfBibitem
\bibitem{LHCb-PROC-2011-006}
M.~Clemencic {\em et~al.}, \ifthenelse{\boolean{articletitles}}{\emph{{The
  \lhcb simulation application, Gauss: Design, evolution and experience}},
  }{}\href{https://doi.org/10.1088/1742-6596/331/3/032023}{J.\ Phys.\ Conf.\
  Ser.\  \textbf{331} (2011) 032023}\relax
\mciteBstWouldAddEndPuncttrue
\mciteSetBstMidEndSepPunct{\mcitedefaultmidpunct}
{\mcitedefaultendpunct}{\mcitedefaultseppunct}\relax
\EndOfBibitem
\bibitem{LHCb-DP-2018-004}
D.~M{\"u}ller, M.~Clemencic, G.~Corti, and M.~Gersabeck,
  \ifthenelse{\boolean{articletitles}}{\emph{{ReDecay: A novel approach to
  speed up the simulation at LHCb}},
  }{}\href{https://doi.org/10.1140/epjc/s10052-018-6469-6}{Eur.\ Phys.\ J.\
  \textbf{C78} (2018) 1009},
  \href{http://arxiv.org/abs/1810.10362}{{\normalfont\ttfamily
  arXiv:1810.10362}}\relax
\mciteBstWouldAddEndPuncttrue
\mciteSetBstMidEndSepPunct{\mcitedefaultmidpunct}
{\mcitedefaultendpunct}{\mcitedefaultseppunct}\relax
\EndOfBibitem
\bibitem{LHCb-DP-2018-001}
R.~Aaij {\em et~al.}, \ifthenelse{\boolean{articletitles}}{\emph{{Selection and
  processing of calibration samples to measure the particle identification
  performance of the LHCb experiment in Run 2}},
  }{}\href{https://doi.org/10.1140/epjti/s40485-019-0050-z}{Eur.\ Phys.\ J.\
  Tech.\ Instr.\  \textbf{6} (2019) 1},
  \href{http://arxiv.org/abs/1803.00824}{{\normalfont\ttfamily
  arXiv:1803.00824}}\relax
\mciteBstWouldAddEndPuncttrue
\mciteSetBstMidEndSepPunct{\mcitedefaultmidpunct}
{\mcitedefaultendpunct}{\mcitedefaultseppunct}\relax
\EndOfBibitem
\bibitem{LHCb-DP-2012-004}
R.~Aaij {\em et~al.}, \ifthenelse{\boolean{articletitles}}{\emph{{The \lhcb
  trigger and its performance in 2011}},
  }{}\href{https://doi.org/10.1088/1748-0221/8/04/P04022}{JINST \textbf{8}
  (2013) P04022}, \href{http://arxiv.org/abs/1211.3055}{{\normalfont\ttfamily
  arXiv:1211.3055}}\relax
\mciteBstWouldAddEndPuncttrue
\mciteSetBstMidEndSepPunct{\mcitedefaultmidpunct}
{\mcitedefaultendpunct}{\mcitedefaultseppunct}\relax
\EndOfBibitem
\bibitem{BBDT}
V.~V. Gligorov and M.~Williams,
  \ifthenelse{\boolean{articletitles}}{\emph{{Efficient, reliable and fast
  high-level triggering using a bonsai boosted decision tree}},
  }{}\href{https://doi.org/10.1088/1748-0221/8/02/P02013}{JINST \textbf{8}
  (2013) P02013}, \href{http://arxiv.org/abs/1210.6861}{{\normalfont\ttfamily
  arXiv:1210.6861}}\relax
\mciteBstWouldAddEndPuncttrue
\mciteSetBstMidEndSepPunct{\mcitedefaultmidpunct}
{\mcitedefaultendpunct}{\mcitedefaultseppunct}\relax
\EndOfBibitem
\bibitem{LHCb-PROC-2015-018}
T.~Likhomanenko {\em et~al.}, \ifthenelse{\boolean{articletitles}}{\emph{{LHCb
  topological trigger reoptimization}},
  }{}\href{https://doi.org/10.1088/1742-6596/664/8/082025}{J.\ Phys.\ Conf.\
  Ser.\  \textbf{664} (2015) 082025},
  \href{http://arxiv.org/abs/1510.00572}{{\normalfont\ttfamily
  arXiv:1510.00572}}\relax
\mciteBstWouldAddEndPuncttrue
\mciteSetBstMidEndSepPunct{\mcitedefaultmidpunct}
{\mcitedefaultendpunct}{\mcitedefaultseppunct}\relax
\EndOfBibitem
\bibitem{PDG2024}
Particle Data Group, N.~S.\ {\em et~al.},
  \ifthenelse{\boolean{articletitles}}{\emph{{\href{http://pdg.lbl.gov/}{Review
  of particle physics}}},
  }{}\href{https://doi.org/10.1103/PhysRevD.110.030001}{to be published in
  Phys.\ Rev \textbf{D110} (2024) 030001}\relax
\mciteBstWouldAddEndPuncttrue
\mciteSetBstMidEndSepPunct{\mcitedefaultmidpunct}
{\mcitedefaultendpunct}{\mcitedefaultseppunct}\relax
\EndOfBibitem
\bibitem{Scikit-learn-paper}
F.~Pedregosa {\em et~al.},
  \ifthenelse{\boolean{articletitles}}{\emph{{Scikit-learn: Machine learning in
  Python}}, }{}J.\ Machine Learning Res.\  \textbf{12} (2011) 2825,
  \href{http://arxiv.org/abs/1201.0490}{{\normalfont\ttfamily
  arXiv:1201.0490}}, {and online at
  \href{http://scikit-learn.org/stable/}{{\texttt{http://scikit-learn.org/stable/}}}}\relax
\mciteBstWouldAddEndPuncttrue
\mciteSetBstMidEndSepPunct{\mcitedefaultmidpunct}
{\mcitedefaultendpunct}{\mcitedefaultseppunct}\relax
\EndOfBibitem
\bibitem{10.1145/307400.307439}
A.~Blum, A.~Kalai, and J.~Langford,
  \ifthenelse{\boolean{articletitles}}{\emph{Beating the hold-out: bounds for
  k-fold and progressive cross-validation}, }{} in {\em Proceedings of the
  Twelfth Annual Conference on Computational Learning Theory},
  \href{https://doi.org/10.1145/307400.307439}{ COLT '99, (New York, NY, USA),
  203–208, Association for Computing Machinery, 1999}\relax
\mciteBstWouldAddEndPuncttrue
\mciteSetBstMidEndSepPunct{\mcitedefaultmidpunct}
{\mcitedefaultendpunct}{\mcitedefaultseppunct}\relax
\EndOfBibitem
\bibitem{LHCb-PAPER-2016-027}
LHCb collaboration, R.~Aaij {\em et~al.},
  \ifthenelse{\boolean{articletitles}}{\emph{{Measurement of the \CP violating
  phase and decay-width difference in \mbox{\decay{\Bs}{\psitwos\phi}}
  decays}}, }{}\href{https://doi.org/10.1016/j.physletb.2016.09.028}{Phys.\
  Lett.\  \textbf{B762} (2016) 253},
  \href{http://arxiv.org/abs/1608.04855}{{\normalfont\ttfamily
  arXiv:1608.04855}}\relax
\mciteBstWouldAddEndPuncttrue
\mciteSetBstMidEndSepPunct{\mcitedefaultmidpunct}
{\mcitedefaultendpunct}{\mcitedefaultseppunct}\relax
\EndOfBibitem
\bibitem{Hulsbergen:2005pu}
W.~D. Hulsbergen, \ifthenelse{\boolean{articletitles}}{\emph{{Decay chain
  fitting with a Kalman filter}},
  }{}\href{https://doi.org/10.1016/j.nima.2005.06.078}{Nucl.\ Instrum.\ Meth.\
  \textbf{A552} (2005) 566},
  \href{http://arxiv.org/abs/physics/0503191}{{\normalfont\ttfamily
  arXiv:physics/0503191}}\relax
\mciteBstWouldAddEndPuncttrue
\mciteSetBstMidEndSepPunct{\mcitedefaultmidpunct}
{\mcitedefaultendpunct}{\mcitedefaultseppunct}\relax
\EndOfBibitem
\bibitem{Pivk:2004ty}
M.~Pivk and F.~R. Le~Diberder,
  \ifthenelse{\boolean{articletitles}}{\emph{{sPlot: A statistical tool to
  unfold data distributions}},
  }{}\href{https://doi.org/10.1016/j.nima.2005.08.106}{Nucl.\ Instrum.\ Meth.\
  \textbf{A555} (2005) 356},
  \href{http://arxiv.org/abs/physics/0402083}{{\normalfont\ttfamily
  arXiv:physics/0402083}}\relax
\mciteBstWouldAddEndPuncttrue
\mciteSetBstMidEndSepPunct{\mcitedefaultmidpunct}
{\mcitedefaultendpunct}{\mcitedefaultseppunct}\relax
\EndOfBibitem
\bibitem{Santos:2013gra}
D.~Mart{\'\i}nez~Santos and F.~Dupertuis,
  \ifthenelse{\boolean{articletitles}}{\emph{{Mass distributions marginalized
  over per-event errors}},
  }{}\href{https://doi.org/10.1016/j.nima.2014.06.081}{Nucl.\ Instrum.\ Meth.\
  \textbf{A764} (2014) 150},
  \href{http://arxiv.org/abs/1312.5000}{{\normalfont\ttfamily
  arXiv:1312.5000}}\relax
\mciteBstWouldAddEndPuncttrue
\mciteSetBstMidEndSepPunct{\mcitedefaultmidpunct}
{\mcitedefaultendpunct}{\mcitedefaultseppunct}\relax
\EndOfBibitem
\bibitem{LHCb-PAPER-2016-039}
LHCb collaboration, R.~Aaij {\em et~al.},
  \ifthenelse{\boolean{articletitles}}{\emph{{New algorithms for identifying
  the flavour of \Bz mesons using pions and protons}},
  }{}\href{https://doi.org/10.1140/epjc/s10052-017-4731-y}{Eur.\ Phys.\ J.\
  \textbf{C77} (2017) 238},
  \href{http://arxiv.org/abs/1610.06019}{{\normalfont\ttfamily
  arXiv:1610.06019}}\relax
\mciteBstWouldAddEndPuncttrue
\mciteSetBstMidEndSepPunct{\mcitedefaultmidpunct}
{\mcitedefaultendpunct}{\mcitedefaultseppunct}\relax
\EndOfBibitem
\bibitem{LHCb-PAPER-2015-056}
LHCb collaboration, R.~Aaij {\em et~al.},
  \ifthenelse{\boolean{articletitles}}{\emph{{A new algorithm for identifying
  the flavour of \Bs mesons at LHCb}},
  }{}\href{https://doi.org/10.1088/1748-0221/11/05/P05010}{JINST \textbf{11}
  (2016) P05010}, \href{http://arxiv.org/abs/1602.07252}{{\normalfont\ttfamily
  arXiv:1602.07252}}\relax
\mciteBstWouldAddEndPuncttrue
\mciteSetBstMidEndSepPunct{\mcitedefaultmidpunct}
{\mcitedefaultendpunct}{\mcitedefaultseppunct}\relax
\EndOfBibitem
\bibitem{LHCb-PAPER-2011-027}
LHCb collaboration, R.~Aaij {\em et~al.},
  \ifthenelse{\boolean{articletitles}}{\emph{{Opposite-side flavour tagging of
  \B mesons at the LHCb experiment}},
  }{}\href{https://doi.org/10.1140/epjc/s10052-012-2022-1}{Eur.\ Phys.\ J.\
  \textbf{C72} (2012) 2022},
  \href{http://arxiv.org/abs/1202.4979}{{\normalfont\ttfamily
  arXiv:1202.4979}}\relax
\mciteBstWouldAddEndPuncttrue
\mciteSetBstMidEndSepPunct{\mcitedefaultmidpunct}
{\mcitedefaultendpunct}{\mcitedefaultseppunct}\relax
\EndOfBibitem
\bibitem{LHCb-PAPER-2015-027}
LHCb collaboration, R.~Aaij {\em et~al.},
  \ifthenelse{\boolean{articletitles}}{\emph{{\B flavour tagging using charm
  decays at the LHCb experiment}},
  }{}\href{https://doi.org/10.1088/1748-0221/10/10/P10005}{JINST \textbf{10}
  (2015) P10005}, \href{http://arxiv.org/abs/1507.07892}{{\normalfont\ttfamily
  arXiv:1507.07892}}\relax
\mciteBstWouldAddEndPuncttrue
\mciteSetBstMidEndSepPunct{\mcitedefaultmidpunct}
{\mcitedefaultendpunct}{\mcitedefaultseppunct}\relax
\EndOfBibitem
\bibitem{Fazzini:2018dyq}
D.~Fazzini, \ifthenelse{\boolean{articletitles}}{\emph{{Flavour Tagging in the
  LHCb experiment}}, }{} in {\em {Proceedings, 6th Large Hadron Collider
  Physics Conference (LHCP 2018): Bologna, Italy, June 4-9, 2018}},
  \href{https://doi.org/10.22323/1.321.0230}{\textbf{LHCP2018} 230, 2018}\relax
\mciteBstWouldAddEndPuncttrue
\mciteSetBstMidEndSepPunct{\mcitedefaultmidpunct}
{\mcitedefaultendpunct}{\mcitedefaultseppunct}\relax
\EndOfBibitem
\bibitem{LHCb-PAPER-2013-065}
LHCb collaboration, R.~Aaij {\em et~al.},
  \ifthenelse{\boolean{articletitles}}{\emph{{Measurements of the \Bp, \Bz, \Bs
  meson and \Lb baryon lifetimes}},
  }{}\href{https://doi.org/10.1007/JHEP04(2014)114}{JHEP \textbf{04} (2014)
  114}, \href{http://arxiv.org/abs/1402.2554}{{\normalfont\ttfamily
  arXiv:1402.2554}}\relax
\mciteBstWouldAddEndPuncttrue
\mciteSetBstMidEndSepPunct{\mcitedefaultmidpunct}
{\mcitedefaultendpunct}{\mcitedefaultseppunct}\relax
\EndOfBibitem
\bibitem{karbach2014decay}
T.~M. Karbach, G.~Raven, and M.~Schiller,
  \ifthenelse{\boolean{articletitles}}{\emph{Decay time integrals in neutral
  meson mixing and their efficient evaluation},
  }{}\href{http://arxiv.org/abs/1407.0748}{{\normalfont\ttfamily
  arXiv:1407.0748}}\relax
\mciteBstWouldAddEndPuncttrue
\mciteSetBstMidEndSepPunct{\mcitedefaultmidpunct}
{\mcitedefaultendpunct}{\mcitedefaultseppunct}\relax
\EndOfBibitem
\bibitem{PDG2022}
Particle Data Group, R.~L. Workman {\em et~al.},
  \ifthenelse{\boolean{articletitles}}{\emph{{\href{http://pdg.lbl.gov/}{Review
  of particle physics}}}, }{}\href{https://doi.org/10.1093/ptep/ptac097}{Prog.\
  Theor.\ Exp.\ Phys.\  \textbf{2022} (2022) 083C01}\relax
\mciteBstWouldAddEndPuncttrue
\mciteSetBstMidEndSepPunct{\mcitedefaultmidpunct}
{\mcitedefaultendpunct}{\mcitedefaultseppunct}\relax
\EndOfBibitem
\bibitem{LHCb-PAPER-2019-036}
LHCb collaboration, R.~Aaij {\em et~al.},
  \ifthenelse{\boolean{articletitles}}{\emph{{Measurement of \CP violation in
  \mbox{\decay{\Bz}{D^{*\pm} D^{\mp}}} decays}},
  }{}\href{https://doi.org/10.1007/JHEP03(2020)147}{JHEP \textbf{03} (2020)
  147}, \href{http://arxiv.org/abs/1912.03723}{{\normalfont\ttfamily
  arXiv:1912.03723}}\relax
\mciteBstWouldAddEndPuncttrue
\mciteSetBstMidEndSepPunct{\mcitedefaultmidpunct}
{\mcitedefaultendpunct}{\mcitedefaultseppunct}\relax
\EndOfBibitem
\bibitem{LHCb-PAPER-2021-005}
LHCb collaboration, R.~Aaij {\em et~al.},
  \ifthenelse{\boolean{articletitles}}{\emph{{Precise determination of the
  $\Bs$-$\Bsb$ oscillation frequency}},
  }{}\href{https://doi.org/10.1038/s41567-021-01394-x}{Nature Physics
  \textbf{18} (2022) 1},
  \href{http://arxiv.org/abs/2104.04421}{{\normalfont\ttfamily
  arXiv:2104.04421}}\relax
\mciteBstWouldAddEndPuncttrue
\mciteSetBstMidEndSepPunct{\mcitedefaultmidpunct}
{\mcitedefaultendpunct}{\mcitedefaultseppunct}\relax
\EndOfBibitem
\bibitem{efron:1979}
B.~Efron, \ifthenelse{\boolean{articletitles}}{\emph{Bootstrap methods: Another
  look at the jackknife},
  }{}\href{https://doi.org/10.1214/aos/1176344552}{Ann.\ Statist.\  \textbf{7}
  (1979) 1}\relax
\mciteBstWouldAddEndPuncttrue
\mciteSetBstMidEndSepPunct{\mcitedefaultmidpunct}
{\mcitedefaultendpunct}{\mcitedefaultseppunct}\relax
\EndOfBibitem
\bibitem{Skwarnicki:1986xj}
T.~Skwarnicki, {\em {A study of the radiative cascade transitions between the
  Upsilon-prime and Upsilon resonances}}, PhD thesis, Institute of Nuclear
  Physics, Krakow, 1986,
  {\href{http://inspirehep.net/record/230779/}{DESY-F31-86-02}}\relax
\mciteBstWouldAddEndPuncttrue
\mciteSetBstMidEndSepPunct{\mcitedefaultmidpunct}
{\mcitedefaultendpunct}{\mcitedefaultseppunct}\relax
\EndOfBibitem
\bibitem{Wilks:1938dza}
S.~S. Wilks, \ifthenelse{\boolean{articletitles}}{\emph{{The large-sample
  distribution of the likelihood ratio for testing composite hypotheses}},
  }{}\href{https://doi.org/10.1214/aoms/1177732360}{Ann.\ Math.\ Stat.\
  \textbf{9} (1938) 60}\relax
\mciteBstWouldAddEndPuncttrue
\mciteSetBstMidEndSepPunct{\mcitedefaultmidpunct}
{\mcitedefaultendpunct}{\mcitedefaultseppunct}\relax
\EndOfBibitem
\bibitem{GammaCombo}
M.~Kenzie {\em et~al.}, \ifthenelse{\boolean{articletitles}}{\emph{{GammaCombo:
  A statistical analysis framework for combining measurements, fitting datasets
  and producing confidence intervals}}, }{}
\newblock
  doi:~\href{https://doi.org/10.5281/zenodo.3371421}{10.5281/zenodo.3371421}\relax
\mciteBstWouldAddEndPuncttrue
\mciteSetBstMidEndSepPunct{\mcitedefaultmidpunct}
{\mcitedefaultendpunct}{\mcitedefaultseppunct}\relax
\EndOfBibitem
\bibitem{LHCb-PAPER-2016-032}
LHCb collaboration, R.~Aaij {\em et~al.},
  \ifthenelse{\boolean{articletitles}}{\emph{{Measurement of the CKM angle
  $\gamma$ from a combination of LHCb results}},
  }{}\href{https://doi.org/10.1007/JHEP12(2016)087}{JHEP \textbf{12} (2016)
  087}, \href{http://arxiv.org/abs/1611.03076}{{\normalfont\ttfamily
  arXiv:1611.03076}}\relax
\mciteBstWouldAddEndPuncttrue
\mciteSetBstMidEndSepPunct{\mcitedefaultmidpunct}
{\mcitedefaultendpunct}{\mcitedefaultseppunct}\relax
\EndOfBibitem
\end{mcitethebibliography}

\newpage
% LHCb collaboration author list
% Data extracted on September 3rd, 2024 at 12:38pm for paper reference LHCb-PAPER-2024-027
\centerline
{\large\bf LHCb collaboration}
\begin
{flushleft}
\small
R.~Aaij$^{37}$\lhcborcid{0000-0003-0533-1952},
A.S.W.~Abdelmotteleb$^{56}$\lhcborcid{0000-0001-7905-0542},
C.~Abellan~Beteta$^{50}$,
F.~Abudin{\'e}n$^{56}$\lhcborcid{0000-0002-6737-3528},
T.~Ackernley$^{60}$\lhcborcid{0000-0002-5951-3498},
A. A. ~Adefisoye$^{68}$\lhcborcid{0000-0003-2448-1550},
B.~Adeva$^{46}$\lhcborcid{0000-0001-9756-3712},
M.~Adinolfi$^{54}$\lhcborcid{0000-0002-1326-1264},
P.~Adlarson$^{81}$\lhcborcid{0000-0001-6280-3851},
C.~Agapopoulou$^{14}$\lhcborcid{0000-0002-2368-0147}
% ...

{\footnotesize \it

% ...
$^{14}$Universit{\'e} Paris-Saclay, CNRS/IN2P3, IJCLab, Orsay, France\\
% ...
$^{37}$Nikhef National Institute for Subatomic Physics, Amsterdam, Netherlands\\
% ...
$^{46}$Instituto Galego de F{\'\i}sica de Altas Enerx{\'\i}as (IGFAE), Universidade de Santiago de Compostela, Santiago de Compostela, Spain\\
% ...
$^{50}$Physik-Institut, Universit{\"a}t Z{\"u}rich, Z{\"u}rich, Switzerland\\
% ...
$^{54}$H.H. Wills Physics Laboratory, University of Bristol, Bristol, United Kingdom\\
% ...
$^{56}$Department of Physics, University of Warwick, Coventry, United Kingdom\\
% ...
$^{59}$School of Physics and Astronomy, University of Glasgow, Glasgow, United Kingdom\\
$^{60}$Oliver Lodge Laboratory, University of Liverpool, Liverpool, United Kingdom\\
% ...
$^{68}$Syracuse University, Syracuse, NY, United States\\
% ...
$^{81}$Department of Physics and Astronomy, Uppsala University, Uppsala, Sweden, associated to $^{59}$\\
% ...
}
\end{flushleft}

\end{document}